\pdfoutput=1
\documentclass[useAMS,usenatbib]{mn2e}
\usepackage{amssymb}
\usepackage{bbm}
\usepackage{txfonts}
\usepackage{psfrag}
\usepackage{graphicx}
\usepackage{natbib}
\usepackage{if then}
\usepackage{longtable}
\usepackage{color}
\usepackage{tikz}
\usetikzlibrary{shapes,arrows}
\usepackage{verbatim}

\def\del#1{{}}

\title{Bayesian physical reconstruction of initial conditions from large scale structure surveys}
\author[Jens Jasche and 
   Benjamin~D.~Wandelt]
       {Jens Jasche$^{1}$ and 
   Benjamin~D.~Wandelt$^{2,1,3,4}$\\
   \\
$^{1}$ CNRS, UMR7095, Institut d'Astrophysique de Paris, F-75014, Paris, France \\
$^{2}$ UPMC Univ Paris 06, UMR7095, Institut d'Astrophysique de Paris, F-75014, Paris, France \\
$^{3}$ Department of Physics, 1110 W Green Street, University of Illinois at Urbana-Champaign, Urbana, IL 61801, USA\\
$^{4}$ Department of Astronomy, 1002 N Gregory Street, University of Illinois at Urbana-Champaign, Urbana, IL 61801, USA
}

\begin{document}
\date{\today}

\pagerange{\pageref{firstpage}--\pageref{lastpage}} \pubyear{2012}

\maketitle

\label{firstpage}

\begin{abstract}
We present a fully probabilistic, physical model of the non-linearly evolved density field, as probed by realistic galaxy surveys. Our model is valid in the linear and mildly non-linear regimes and uses second order Lagrangian perturbation theory to connect the initial conditions with the final density field. Our parameter space consists of the 3D initial density field and our method allows a fully Bayesian exploration of the sets of initial conditions that are consistent with the galaxy distribution sampling the final density field. A natural byproduct of this technique is an optimal non-linear reconstruction of the present density and velocity fields, including a full propagation of the observational uncertainties. A test of these methods on simulated data mimicking the survey mask, selection function and galaxy number of the SDSS DR7 main sample shows that this physical model gives accurate reconstructions of the underlying present-day density and velocity fields on scales larger than $\sim 6$ Mpc/h.  Our method naturally and accurately reconstructs non-linear features corresponding to three-point and higher order correlation functions such as walls and filaments. Simple tests of the reconstructed initial conditions show statistical consistency with the Gaussian simulation inputs. Our test demonstrates that statistical approaches based on physical models of the large scale structure distribution are now becoming feasible for realistic current and future surveys. \end{abstract}


\section{Introduction and Motivation}

Ongoing and planned Large Scale Structure (LSS) surveys will measure the distribution of galaxies at an unprecendented level of accuracy in the coming decade. These surveys are expected to vastly enhance our constraints on the physics of cosmogenesis, neutrino physics, and dark energy phenomenology.

How do we compare cosmological models to these surveys?
We have an observationally well-supported physical model of the initial conditions. According to this model, a homogeneous and isotropic density field with small, very nearly Gaussian, and nearly scale-invariant correlated density perturbations arose from quantum perturbations in the very early Universe. Gravitational evolution in an expanding background processed these initial conditions into an evolved density field, at first through linear transfer and then through non-linear structure formation. LSS surveys catalogue the positions of observed tracers of this evolved density field in redshift space.

It is now standard to model the initial Gaussian density perturbations statistically in terms of the early universe processes that created them, such as the physics of inflation , the change from matter to radiation dominated universe, neutrino free-streaming, and the acoustic oscillations of photon-baryon plasma. Within the standard cosmology, the evolution and growth of the initial perturbations in an expanding Universe is well-understood in principle, and directly linked to its dominant constituents such as dark matter and dark energy. It therefore seems natural to analyse LSS surveys directly in terms of the simultaneous constraints they place on the initial density field and the physical evolution that links the initial density field to the observed tracers of the evolved density field.

For a variety of good reasons the current state of the art of statistical analyses of LSS surveys is far removed from this ideal. There are some areas where significant progress seems very difficult. In particular, a detailed physical model of the way galaxies arise in response to the spatial fluctuations in the dark matter distribution is not computationally tractable (the ``bias'' problem). Even for the dark matter alone, reversing the non-linear evolution that link the initial and evolved density field is a fundamentally ill-posed problem \citep[see e.g.][]{NUSSER1992,Crocce2006}. 

As a consequence, the state of the art in the analysis of galaxy surveys addresses these problems in isolation.
In the standard approach, the link between theory and observation is made through the power spectrum. This requires solving two separate problems: the data analysis problem of inferring the power spectrum from an observed sample of tracers given a survey mask and selection function \citep[see e.g.][]{FKP,TEGMARK_2004,WANDELT2004,2004ApJS..155..227E,PERCIVAL2005,JASCHE2010PSPEC,ELSNER2012}; and the much more difficult theoretical problem of modeling the power spectrum \textit{and the form of its likelihood} for the non-linearly evolved and biased galaxy density field \citep[see e.g.][and references therein]{BAUGH1995,PEACOCK1996,SMITH2003,JEONG2006,HAITMANN2010}. 

Three-dimensional inference of the matter distribution from observations
requires modeling the statistical behavior of the mildly non-linear and non-linear regime of the matter distribution.
The exact statistical behavior of the matter distribution in terms of a probability distribution for the fully evolved density field is not known. 
Previous approaches therefore relied on phenomenological approximations such as multivariate Gaussian or log-normal distributions incorporating
a cosmological power-spectrum to accurately account for the correct two-point statistics of the density fields. Both of these distributions can be considered
as maximum entropy prior on a linear and logarithmic scale, respectively, and are therefore  well-justified for Bayesian analysis. However, these priors
only parametrize the two-point statistics of the  matter distribution.
Since large scale structure formation through gravitational clustering is essentially a deterministic process described by Einstein's equations and since the only
stochasticity in the problem enters in the generation of initial conditions, it seems reasonable to account for the increasing statistical complexity of the evolving matter distribution
by a dynamical model.

In this paper we describe progress towards such an  approach that uses data to constrain a set of \textit{a priori} possible dynamical, three-dimensional histories. We use second order Lagrangian perturbation theory (2LPT) as a physical model of the gravitational dynamics that link the initial three-dimensional Gaussian density field to the observed, non-Gaussian density field. In Bayesian parlance our prior for the evolved density is the initial Gaussian density field evolved by a 2LPT model. Using the powerful sampling techniques  recently developed by \cite{JASCHE2010HADESMETHOD} we can use this model as prior information and explore the range of initial Gaussian density fields that are statistically consistent with the data, modeled as a Poisson sample from evolved density fields. 

Our method will also automatically generate reconstructions of the large scale velocity field since our model incorporates dynamics. Since the approach is implemented in a fully Bayesian framework we do not produce unique reconstructions, but a set of \textit{samples}  which can be interpreted as a probabilistic representation of  the information the observations contain about the underlying density (initial and evolved) and the velocity field. In particular, the variations between  samples represent the uncertainties that remain in the reconstruction owing to the modeled statistical and systematic errors in the data.

\subsection{Comparison to prior work}
In the recent past several papers have pointed out the promise of the lognormal model in fitting to observations of the non-linear density field \citep[see e.g.][]{KITAURA2010MNRAS,JASCHE2010HADESMETHOD,JASCHE2010HADESDATA}. While the lognormal approach provides a good model of the 1-point and 2-point functions of the field we will show that Gaussian statistics evolved by 2LPT reproduces those successes but, in addition,  reproduces features naturally that are associated with the higher order n-point functions such as filaments and walls. This is not surprising since it is well known that 2LPT  reproduces the exact one- and three-point statistics of fully non-linear density fields at large scales, and also approximates higher order statistics very well \citep[see e.g.][]{MOUTARDE1991,BUCHERT1994,BOUCHET1995,SCOCCIMARRO2000,BERNADEAU2002,PTHALOS}. 

The field of velocity field reconstructions has a long history \citep[see e.g.][]{BERTSCHINGER1990,NUSSER1992,DEKEL1999,FRISCH2002,BRENIER2003,MOHAYAEE2008,LAVAUXMAK2008,KITAURA2011VELOCITY}. The contribution of our approach is the  imbedding of a non-Gaussian model in a probabilistic framework. Zel'dovich and MAK are, respectively, perturbative and non-perturbative attempts to reconstruct the displacement field linking the initial conditions  from tracers of large scale structure and as such also generate estimates of the velocity field. Our approach goes beyond these works in several ways: we combine the inference with a detailed non-Gaussian model of realistic survey features (mask, selection function and shot noise); we implement explicitly a Gaussian prior for the initial density field; and the Bayesian exploration gives a quantitative characterization of the uncertainties in our inferences.

Significant effort has also been invested in establishing accurate representations of the observed Universe in numerical simulations, by constraining simulations by observations \citep[see e.g.][]{KRAVTSOV2002,KLYPIN2003,DOLAG2005,LIBESKIND2010,GOTTLOEBER2009,GOTTLOEBER2010,LAVAUX2010}.
Many of these approaches rely on  integrating the observed present day density field backwards in time to the initial state. Such an approach is generally hindered due to incomplete observations
of the final state and by spurious erronous enhancement of decaying mode power in the initial conditions during backward integration \citep[][]{NUSSER1992}.
The fully probabalistic approach, proposed in this work, naturally accounts for uncertainties of only partially observed final states, by exploring
physical reasonable solutions, filtered by the 2LPT model, for the initial conditions which can all lead to the same or similar final observations.
Furthermore, our method solely depends on forward evaluations of the model, which therefore accurately handels the issue of decaying mode power.
Also note that  unique recovery of initial conditions is generally not possible on all scales due to the chaotic nature of the 
dynamical large scale structure formation process on small scales \citep[see e.g.][]{NUSSER1992,Crocce2006}. These uncertainties  will also be accurately
accounted for by our method while exploiting information on the initial conditions on all scales accessible to the 2LPT model.

The paper is structured as follows. In section \ref{2lpt_posterior} we discuss the design of posterior distributions for large scale structure 
inference and show that the complex problem of modeling accurate prior distributions for the evolved non-Gaussian matter distribution can be recast as an initial conditions problem once a physical model for large scale structure
formation is specified. Furthermore, we will present the resultant 2LPT-Poissonian posterior distribution for the inference of the three dimensional matter distribution
from galaxy surveys. Section \ref{HAMILTONIAN_SAMPLING} provides a brief overview over the Hamiltonian sampling approach employed in the inference framework described in this work, and in section \ref{equations_of_motion} we present the relevant derivations of the Hamiltonian forces required for an
efficient numerical implementation of the Hybrid Monte Carlo sampler. In the following section \ref{mock_observations}, we describe the generation of an artificial galaxy survey, inspired by 
 the Sloan Digital Sky Survey data release 7 main sample \citep[][]{SDSS7}. In section \ref{testing} we describe the application of our method
to this simulated  data in order to provide a proof of concept and to estimate the behavior of the algorithm in a realistic setting.
We will conclude the paper with a summary and a discussion of the results in section \ref{Conclusion}.

\section{The 2LPT-Poissonian posterior}
\label{2lpt_posterior}
\subsubsection{The non-Gaussian density prior}
\label{density_prior}
As already pointed out in the introduction, inferring the three dimensional large scale structure from observations requires the design of suitable prior distributions for the fully gravitationally evolved density field.
Standard approaches such as Wiener filtering employ Gaussian priors, which are assumed to be suitable for the inference of the largest scales\citep[see e.g.][]{LAHAV1994,ZAROUBI2002,ERDOGDU2004,KITAURA2008MNRAS,KITAURA2009MNRAS,JASCHE2010PSPEC}.  
For the inference of the density field in the non-linear regime log-normal priors have been proposed and successfully applied to large scale structure inference problems \citep[][]{KITAURA2010MNRAS,JASCHE2010HADESMETHOD,JASCHE2010HADESDATA}.
More recently, \citet{KITAURA2012} proposed to use Edgeworth expansions to construct prior distributions incorporating also third order moments of the distribution.
All of these approaches are based on heuristic approximations to the full problem. Currently, a closed form description of the present day density field in terms of a multivariate probability distribution does not exist.

While there exist considerable difficulties in constructing a suitable probability distribution for the present day density field, the initial seed fluctuations at redshifts \(z\sim 1000\)
 obey Gaussian statistics to great accuracy, in agreement with theories of inflation and observations \citep[see e.g. ][]{LINDE2008,KOMATSU2011}.
Therefore, the complicated nature of the present matter distribution solely originates from deterministic physical processes during structure formation.
Generally, gravitational interactions introduce mode coupling and phase correlations, such that the statistical behavior of the present day density  
 strongly deviate from a Gaussian distribution \citep[see e.g.][]{Peacockbook}.

Since initial and final conditions are linked via deterministic structure formation processes, it seems reasonable to formulate the inference problem in 
terms of simpler statistics at the initial conditions, rather than approximating the complex statistical behavior of the non-linear matter distribution.
More specifically, given a suitable model of large scale structure formation \(G(a,\delta^i)\) we can obtain a prior distribution for the final density contrast \(\delta^f\)  for a given cosmic scale factor \(a\) by marginalizing over the initial conditions:
\begin{eqnarray}
{\mathcal P}(\{\delta^f_l\}) &=& \int \mathrm{d}\{\delta^i_l\}\, {\mathcal P}(\{\delta^f_l\},\{\delta^i_l\})\, \nonumber \\ 
&=& \int \mathrm{d}\{\delta^i_l\}\, {\mathcal P}(\{\delta^i_l\})\, {\mathcal P}(\{\delta^f_l\}|\{\delta^i_l\})\, , \nonumber \\
\end{eqnarray}
where, for a deterministic structure formation model, the conditional probability is given by Dirac delta distributions:
\begin{equation}
{\mathcal P}(\{\delta^f_l\}|\{\delta^i_l\})) = \prod_l \delta^D(\delta^f_l-G(a,\delta^i)_l)\, .
\end{equation}
Given a model \(G(a,\delta^i)\) for structure formation, a prior distribution for the present day density field can be obtained by a two step sampling process, by
first generating an initial conditions realization from the prior distribution \({\mathcal P}(\{\delta^i_l\})\) and then propagating the initial state forward in time
with a suitable model of large scale structure formation.
This process amounts to generating samples from the joint prior distribution of the final and initial conditions:
\begin{equation}
{\mathcal P}(\{\delta^f_l\},\{\delta^i_l\}) = {\mathcal P}(\{\delta^i_l\})\, \prod_l \delta^D(\delta^f_l-G(a,\delta^i)_l)\, .
\end{equation}
By discarding the initial density realization, one obtains a realization from the prior distribution for the non-linear
final state.
Assuming, a multivariate Gaussian process with zero mean and covariance matrix \(S\) to generate the initial density field \(\delta^i\) the joint prior distribution
is given as: 
\begin{eqnarray}
\label{joint_density_prior}
{\mathcal P}(\{\delta^i_l\},\{\delta^f_l\}|S)) &=& {\mathcal P}(\{\delta^i_l\}|S)\, \prod_l \delta^D\left(\delta^f_l-G(a,\delta^i)_l\right)\, \nonumber \\
 &=& \frac{\mathrm{e}^{ -\frac{1}{2}\,\sum_{lm} \delta^i_l S_{lm}^{-1} \delta^i_m}}{\mathrm{det}\left(2\, \pi S \right)} \, \prod_l \delta^D\left(\delta^f_l-G(a,\delta^i)_l\right)\, .\nonumber \\
\end{eqnarray}

In this work, we will employ a second order Lagrangian perturbation theory (2LPT) model to approximately describe gravitational large scale structure formation (also see appendix \ref{2lptmodel} for an overview over the 2LPT model).
2LPT is able to recover the exact one-, two- and three-point statistics at large scales, and also approximates higher order statistics very well \citep[see e.g.][]{MOUTARDE1991,BUCHERT1994,BOUCHET1995,SCOCCIMARRO2000,PTHALOS}.
The 2LPT model therefore naturally reproduces physically reasonable higher order statistics in the matter inference problem without requiring
the introduction of additional parameters for the description of higher order statistics.
Our approach therefore provides a physically meaningful way of  matching the higher order statistics of the evolved density field while requiring no further knowledge other
than the initial two-point statistics.
 
\subsubsection{The large scale structure likelihood}
\label{likelihood}
Above we demonstrated that  the task of modeling accurate prior distributions for the statistical behavior of the present day matter distribution
can be recast into an initial conditions inference problem once a model for large scale structure formation is specified.

The methods described in this work are general and can be adapted for the inference from any particular probe of the three dimensional large scale structure.
We will illustrate our method for the case of optical galaxy redshift surveys.

Galaxies tend to follow the gravitational potential of the cosmic matter distribution and thus can be considered as matter tracers.
The statistical uncertainty due to the discrete nature of the galaxy distribution can be modeled as an inhomogeneous Poisson processe \citep[see e.g.][]{LAYZER1956,PEEBLES1980,MARTINEZ2002}.
Also note that  Poissonian likelihoods have already been successfully employed for non-linear density field inference \citep[see e.g.][ for details]{KITAURA2010MNRAS,JASCHE2010HADESMETHOD,JASCHE2010HADESDATA}.
Following this approach, the corresponding Poissonian likelihood distribution can be expressed as:
\begin{equation}
\label{eq:Poissonian}
{\mathcal P}(\{N_k^{g}\}|\{\lambda_k\})= \prod_k \frac{{\left(\lambda_k\right)}^{N^{g}_k} e^{-\lambda_k}}{{N^{g}_k}!} \, ,
\end{equation}
where \(N_k^{g}\) is the observed galaxy number at position \(\vec{x}_k\) in the sky and  \(\lambda_k\) is the expected number of galaxies at this position.
The mean galaxy number is related to the final density field \(\delta^f_k\) via:
\begin{equation}
\label{eq:data_model}
\lambda_k = \lambda_k\left(\delta\right)= R_k \bar{N}(1+B(\delta^f)_k)\, ,
\end{equation}
where \(R_k\) is a linear response operator, incorporating survey geometries and selection effects, \(\bar{N}\) is the mean number of galaxies in the volume and \(B(x)_k\) is a nonlinear, non local, bias operator at position \(\vec{x}_k\) \citep[also see ][ for further discussions]{JASCHE2010HADESMETHOD,JASCHE2010HADESDATA}.

The joint posterior distribution for the initial conditions \(\delta^i_l\) and the final density field \(\delta^f_l\) given the galaxy observations is then obtained by the multiplying equation (\ref{joint_density_prior}) and (\ref{eq:Poissonian}):
\begin{eqnarray}
\label{eq:2lpt_posterior}
{\mathcal P}(\{\delta^i_l\},\{\delta^f_l\}|\{N_i\},S) &=&   \frac{\mathrm{e}^{ -\frac{1}{2}\,\sum_{lm} \delta^i_l S_{lm}^{-1} \delta^i_m}}{\mathrm{det}\left(2\, \pi S \right)} \, \prod_l \delta^D\left(\delta^f_l-G(a,\delta^i)_l\right)\, \nonumber \\ \nonumber \\
& &  \times\, \prod_k \frac{{\left(\lambda_k\left (\delta^f \right)\right)}^{N^{g}_k} e^{-\lambda_k\left( \delta^f \right)}}{{N^{g}_k}!}\, .
\end{eqnarray}
We see that given a model of structure formation \(G(a,\delta^i)\), the final density field \(\delta^f_l\) is a free byproduct of the inference process.
Consequently, marginalizing equation (\ref{eq:2lpt_posterior}) over \(\delta^f_l\) yields the desired target posterior distribution for large scale structure inference:
\begin{eqnarray}
\label{eq:final_2lpt_posterior}
{\mathcal P}\left(\{\delta^i_l\}|\{N_i\},S\right) &=&   \frac{\mathrm{e}^{ -\frac{1}{2}\,\sum_{lm} \delta^i_l S_{lm}^{-1} \delta^i_m}}{\mathrm{det}\left(2\, \pi S \right)} \,  \nonumber \\ \nonumber \\
& &  \times\, \prod_k \frac{{\left(\lambda_k\left (G(a,\delta^i) \right)\right)}^{N^{g}_k} e^{-\lambda_k\left( G(a,\delta^i) \right)}}{{N^{g}_k}!}\, .
\end{eqnarray}
This result requires several remarks.
First, A nearly trivial, but nevertheless important, conclusion to draw from (\ref{eq:final_2lpt_posterior}) is that large scale structure inference
depends solely on the initial conditions \(\delta^i_l\). Consequently, the complex task of designing
suitable prior distributions for the inference of the evolved density field can be recast into an initial value problem by modeling a suitable 
physical model to account for the complexity of the final state. 

Second, note that inferring the initial density field does not involve backward in time integration of the physical model. 
The inference process exclusively requires model evaluations in the forward time direction, counter to the widely held notion that inference of initial conditions requires backward integration of the equations of motion.
Nevertheless, traditional approaches of initial conditions inference, also known as 'time machines', rely on the inversion of the flow of time in the model equations \citep[see e.g.][]{NUSSER1992}.
As pointed out by \citet{NUSSER1992}, the disadvantage of backward integration is that it may lead to artificial increase of decaying modes amplitudes introducing erroneous artificial density and velocity fluctuations in the initial conditions.
Also note that  large scale structure surveys only provide limited information on the full final state, due to survey geometries and statistical uncertainties. These problems generally hinder a unique backward integration of the partially observed final state to the initial conditions.

To alleviate this problem, and to ensure physical meaningful backward integration of non-linear models, one has to augment unobserved regions in the data 
with statistically meaningful information mimicking the unobserved part of the evolved density field. This in turn requires accurate knowledge on the multivariate
probability distribution for the evolved final state \(\delta^f_l\), which is not known at present.

Such problems are naturally prevented by casting the reconstruction of initial conditions as the statistical inference problem  expressed in equation (\ref{eq:final_2lpt_posterior}).
Since the proposed method solely depends on forward model evaluations, unobserved regions and statistical uncertainties 
can be easily dealt with in the initial conditions, which amounts to modeling simple, uncorrelated Gaussian processes.
Further, the posterior distribution proposed in equation
(\ref{eq:final_2lpt_posterior}) accounts for systematics, such as survey geometry, selection effects and biases but also for statistical uncertainties such as the noise of the galaxy distribution and cosmic variance.  

We therefore see that  statistical uncertainties do not allow a unique inference of the initial conditions from partially observed final states.
Consequently, our approach aims at exploring the highly non-Gaussian and non-linear posterior distribution \({\mathcal P}\left(\{\delta^i_l\}|\{N_i\},S\right)\) of the initial density field \(\delta^i_l\) conditional on galaxy observations
\(N_l\) in order to quantify the uncertainty and significance of the inferred initial conditions.
The overall inference process is numerically non-trivial. It is made possible by sophisticated non-linear analysis methods, which will be described in the following.

\begin{figure*}
\centering{\includegraphics{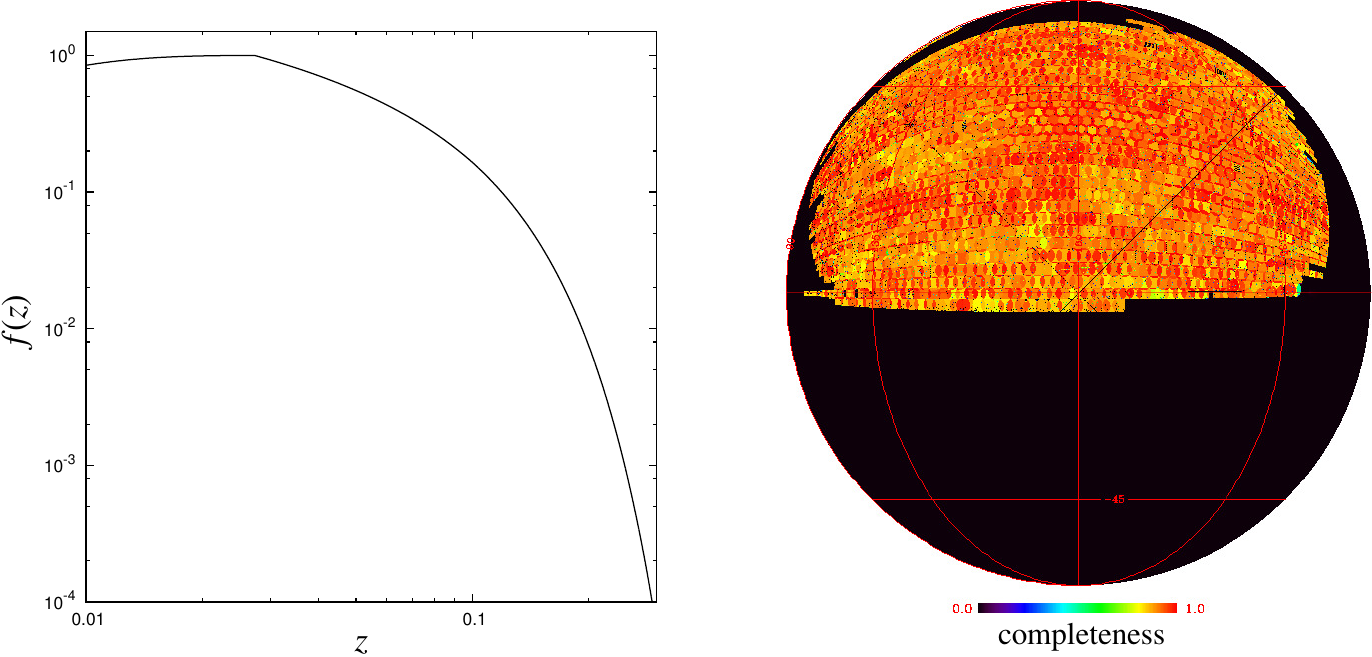}}
\caption{Selection function \(f(z)\) as a function of redshift \(z\) (left panels) and the two dimensional completeness map for the SDSS DR7 (right panel).}
\label{fig:TEST_SEL_WIN}
\end{figure*}

\section{Hamiltonian sampling}
\label{HAMILTONIAN_SAMPLING}
As described in the previous section, exploration of the initial conditions posterior distribution requires numerically efficient Markov Chain Monte Carlo methods to accurately
account for all non-linearities and non-Gaussianities involved in the inference process.
Unfortunately, unlike as in the Gibbs sampling approach for large scale structure proposed in \citet{JASCHE2010PSPEC}, direct sampling from this posterior is not possible.
One therefore has to rely on a sampling procedures with an accept-reject step for the exploration of the high dimensional parameter space encountered in this problem.
A major draw back of traditional algorithms of this type, e.g. Metropolis-Hastings, is their dominant random walk behavior and a possible high rejection rate if no suitable proposal distribution can be designed.
In this work, we usually deal with about \(10^6\), or more, free parameters \(\delta^i_l\) which correspond to the initial density contrast amplitudes at the volume elements of the analyzed volume. 
Due to this high dimensionality of the problem, a low acceptance rate of the Metropolis-Hastings algorithm would result in a prohibitive computational cost for the method.
Given this situation, we propose to use a Hybrid Monte Carlo (HMC) method, which in the absence of numerical errors, would yield an acceptance rate of unity.
The HMC method exploits techniques developed to follow classical dynamical particle motion in potentials \citep[][]{DUANE1987,NEAL1993,NEAL1996}. In this fashion the Markov sampler follows a persistent motion through the parameter space, suppressing the random walk behavior.
This enables us to sample with reasonable efficiency in high dimensional spaces \citep[][]{HANSON2001}.
Furthermore, the HMC has already been successfully applied to non-linear large scale structure inference problems \citep[see e.g.][]{JASCHE2010HADESMETHOD,JASCHE2010HADESDATA}.

In the following, we will just briefly outline the basic idea of the hybrid Hamiltonian sampling algorithm. More detailed description of the algorithm and its application in large scale structure inference and in general can be found in \citep[][]{DUANE1987,NEAL1993,HANSON2001,JASCHE2010HADESMETHOD,JASCHE2010HADESDATA}. 

\subsection{The HMC}

Suppose, we wish to generate samples from a probability distribution \({\mathcal P}(\{x_i\})\), where \(\{x_i\}\) is a set consisting of the \(N\) elements \(x_i\). If we interpret the negative logarithm of this posterior distribution as a potential:
\begin{equation}
\label{eq:Potential}
\psi(x)=-ln({\mathcal P}(x)) \, ,
\end{equation}
and by introducing a 'momentum' variable \(p_i\) and a 'mass matrix' \(M\), as nuisance parameters, we can formulate a Hamiltonian describing the dynamics in the multi dimensional phase space.
Such a Hamiltonian is then given as:
\begin{equation}
\label{eq:Hamiltonian}
H = \sum_i\sum_j \frac{1}{2}\,p_i\,M_{ij}^{-1}\,p_j +\psi(x) \, .
\end{equation}
As can be seen in equation (\ref{eq:Hamiltonian}), the form of the Hamiltonian is such that  the joint distribution is separable into a Gaussian distribution in the momenta \(\{p_i\}\) and the target distribution \({\mathcal P}(\{x_i\})\) as:
\begin{equation}
\label{eq:TARGET_DISTRIBUTION}
e^{-H} = {\mathcal P}(\{x_i\})\,e^{-\frac{1}{2}\,\sum_i\sum_j\,p_i\,M_{ij}^{-1}\,p_j}\, .
\end{equation}
For this reason, marginalization over all momenta will again yield the original target distribution \({\mathcal P}(\{x_i\})\).

In order to generate a random variate from this joint distribution, being proportional to \(\exp(-H)\), one first draws a set of momenta from the distribution defined by the kinetic energy term that  is an \(N\) dimensional Gaussian with a covariance matrix \(M\).
Then  one follows the deterministic dynamical evolution of the system, given a starting point \( (\{x_i\},\{p_i\})\) in phase space for some fixed pseudo time \(\tau\) according to Hamilton's equations:
\begin{equation}
\label{eq:HAMILTON1}
\frac{dx_i}{dt} = \frac{\partial H}{\partial p_i}\, .
\end{equation}

\begin{equation}
\label{eq:HAMILTON2}
\frac{dp_i}{dt} = \frac{\partial H}{\partial x_i} = - \frac{\partial \psi(x)}{\partial x_i}\, .
\end{equation}
The integration of this equations of motion yields the new position \((\{x'_i\},\{p'_i\})\) in phase space. This new point is accepted according to the usual acceptance rule:
\begin{equation}
\label{eq:acceptance_rule}
{\mathcal P}_A = min\left[1,\exp(-\left(H(\{x'_i\},\{p'_i\})-H(\{x_i\},\{p_i\})\right)\right]\, .
\end{equation}
Since the equations of motion provide a solution to a Hamiltonian system, energy or the Hamiltonian given in equation (\ref{eq:Hamiltonian}) is conserved, and therefore the solution to this system provides an acceptance rate of unity. In practice, numerical errors can lead to a somewhat lower acceptance rate. 
Samples of the desired target distribution are then obtained by simply discarding the auxiliary momenta \(\{p_i\}\), which amounts to marginalization over these nuisance parameters.
The particular implementation of the hybrid Hamiltonian Monte Carlo sampler for the problem described in this work will be discussed below.

\begin{figure*}
\centering{\includegraphics[width=0.4\textwidth,clip=true]{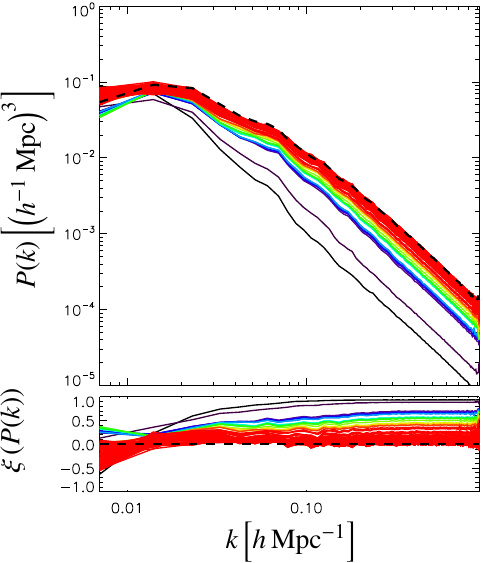}}
\caption{The plot demonstrates the initial burn-in drift of successive power-spectra, measured from the initial density fields, towards the true underlying solution. Successive samples are color coded corresponding to their sample number as indicated by the color bar on the right. Black dashed lines correspond to the true underlying values.
Lower panels depict the successive deviation \(\xi\) from the true values, as described in the text, for the measured power-spectra.
The sequence of 800 successive samples, visualizes how the sampler approaches the true underlying values and starts exploring the parameter space around them.}
\label{fig:burn_in}
\end{figure*}

\section{Equations of motion for LSS inference}
\label{equations_of_motion}
As described above, the HMC approach permits to explore the non-linear large scale structure posterior by following Hamiltonian dynamics in the high dimensional parameter space. 
The corresponding forces, required to evaluate these Hamiltonian trajectories can be derived from the large scale structure posterior given in equation (\ref{eq:final_2lpt_posterior}).
Consequently, the Hamiltonian potential \(\Psi(\{\delta^i_l\})\) can be written as:
\begin{eqnarray}
\label{eq:density_post}
\Psi(\{\delta^i_l\}) &=& -{\rm{ln}}\left({\mathcal P}(\{\delta^i_l\}|\{N_i\},S)\right)\nonumber \\
&=& \Psi_{prior}(\{\delta^i_l\}) + \Psi_{likelihood}(\{\delta^i_l\}) \, ,\nonumber \\
\end{eqnarray}
with the potential \(\Psi_{prior}(\{\delta^i_l\})\) is given as:
\begin{eqnarray}
\Psi_{prior}(\{\delta^i_l\}) &=&  \frac{1}{2} \sum_{lm} \delta^i_l S^{-1}_{lm}\, \delta^i_m \, ,
\end{eqnarray}
and \(\Psi_{likelihood}(\{\delta^i_l\})\) is given as:
\begin{eqnarray}
\Psi_{likelihood}(\{\delta^i_l\}) &=& \sum_k R_k\bar{N}_{gal}\,(1+G(a,\delta^i)_k) \nonumber \\
& &- N_k {\rm{ln}}\left(R_k\bar{N}_{gal}\,(1+G(a,\delta^i)_k)\right) \, ,
\end{eqnarray}
Given the above definition of the Hamiltonian potential \(\Psi(\{\delta^i_l\})\) one can obtain the required Hamiltonian forces by differentiating with respect to \(\delta^i_p\): 
\begin{eqnarray}
\label{eq:Hamiltonian_forces}
\frac{\partial\Psi(\{\delta^i_l\})}{\partial \delta^i_p} &=& \frac{\partial\Psi_{prior}(\{\delta^i_l\})}{\partial \delta^i_p} + \frac{\partial\Psi_{likelihood}(\{\delta^i_l\})}{\partial \delta^i_p}\, ,
\end{eqnarray}
Here the prior term is given by:
\begin{eqnarray}
\label{eq:prior_force}
\frac{\partial\Psi_{prior}(\{\delta^i_l\})}{\partial \delta^i_p}  &=& \sum_{j} S^{-1}_{pj}\,\delta^i_j \, .
\end{eqnarray}
In contrast the likelihood term cannot be derived as trivially. A detailed derivation for the likelihood term can be found in Appendix \ref{Ap:Hamiltonian_force}.
The likelihood term \(\Psi_{likelihood}(\{\delta^i_l\}))\) can be expressed as:
\begin{eqnarray}
\label{eq:likelihood_force}
\frac{\partial\Psi_{likelihood}(\{\delta^i_l\})}{\partial \delta^i_p}  &=& - D^1\, J_p + D^2 \sum_{a>b} \left( \tau_p^{aabb} + \tau_p^{bbaa} - 2 \tau_p^{abab} \right) \, ,\nonumber \\
\end{eqnarray}
where the vector \(J_p\) and the tensor \(\tau_p^{abcd}\) are defined in Appendix \ref{Ap:Hamiltonian_force}.

Finally, the equations of motion for the Hamiltonian system given in equations (\ref{eq:HAMILTON1}) and (\ref{eq:HAMILTON2}) can be written as:

\begin{equation}
\label{eq:HAMILTON1_2lpt}
\frac{d\delta^i_n}{dt} = \sum_{j} M^{-1}_{nj}\,p_j\, ,
\end{equation}
and
\begin{equation}
\label{eq:HAMILTON2_2lpt}
\frac{dp_n}{dt} = - \sum_{j} S^{-1}_{nj}\,\delta^i_j + D^1\, J_n + D^2 \sum_{a>b} \left( \tau_n^{aabb} - \tau_n^{bbaa} - 2 \tau_n^{abab} \right)  \, .
\end{equation}
New samples from the large scale structure posterior can then be obtained by following the dynamical evolution of the Hamiltonian system in phase space.

\section{Numerical Implementation}
\label{Numerical_implementation}
We named our numerical implementation of the initial conditions sampler BORG (Bayesian Origin Reconstruction from Galaxies).
It utilizes the FFTW3 library for Fast Fourier Transforms and the GNU scientific library (gsl) for random number generation \citep{FFTW05,GSL}.
In particular, we use the Mersenne Twister MT19937, with 32-bit word length, as provided by the gsl\_rng\_mt19937 routine, which was particularly designed for Markov Chain Monte Carlo simulations \citep{MERSENNE_TWISTER}.

\subsection{The leapfrog scheme}
\label{LEAPFROG}
The numerical implementation of the HMC sampler employed in this work largely follows the implementation described in \cite{JASCHE2010HADESMETHOD}.
Generally the numerical integration scheme  is required to meet some conditions in order to achieve optimal efficiency of the sampler. 
High acceptance rates require the numerical integration scheme to be highly accurate in order to conserve the total Hamiltonian. Low accuracy in the integration scheme will generally lower the acceptance rate. Additionally, the integrator must be symplectic, meaning exactly reversible, in order to ensure the Markov Chain satisfies detailed balance \citep{DUANE1987}.
For this reason, we implemented the leapfrog scheme to integrate the Hamiltonian system. It is also numerically robust, and allows for simple propagation of errors. 
In particular, we implement the Kick-Drift-Kick scheme.
The equations of motions are integrated by making \(n\) steps with a finite step size \(\epsilon\), such that \(\tau=n  \epsilon\):

\begin{equation}
\label{eq:LEAPFROG1}
p_m\left(t+\frac{\epsilon}{2}\right) = p_m(t) -\frac{\epsilon}{2} \left .\frac{\partial  \psi(\{\delta^i_k\})}{\partial \delta^i_l} \right |_{\delta^i_m(t)} \, , 
\end{equation}

\begin{equation}
\label{eq:LEAPFROG2}
\delta^i_m\left(t+\epsilon \right) = \delta^i_m(t) -\frac{\epsilon}{m_i}\,p_m\left(t+\frac{\epsilon}{2}\right)  \, , 
\end{equation}

\begin{equation}
\label{eq:LEAPFROG3}
p_m\left(t+\epsilon\right) = p_m\left(t+\frac{\epsilon}{2}\right) -\frac{\epsilon}{2} \left .\frac{\partial  \psi(\{\delta^i_k\})}{\partial \delta^i_l} \right |_{\delta^i_m\left(t+\epsilon \right)} \, .  
\end{equation}
We iterate these equations until \(t=\tau\).
Also note that  it is important to vary the pseudo time interval \(\tau\), to avoid resonant trajectories. We do so by drawing \(n\) and \(\epsilon\) randomly from a uniform distribution.

\begin{figure*}
\centering{\includegraphics[width=1.0\textwidth,clip=true]{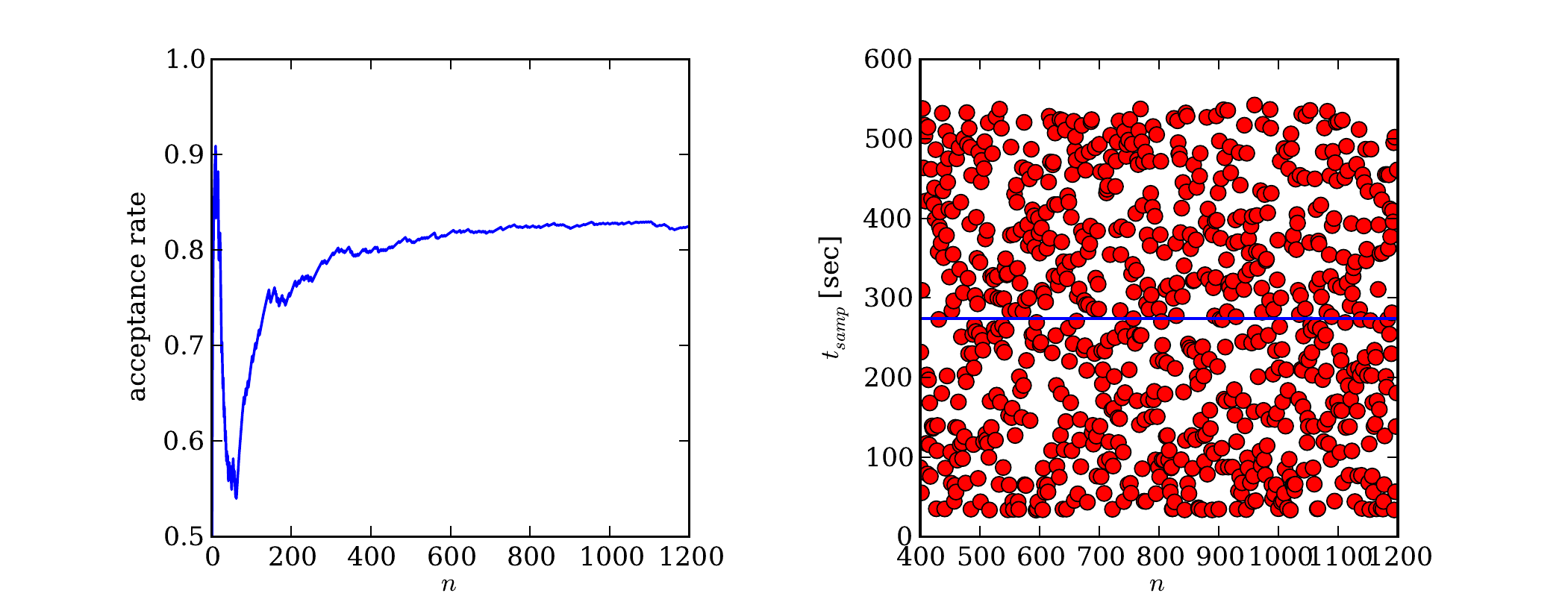}}
\caption{Acceptance rates for successive samples (left panel) and the execution time per sample (right panel). It can be seen that the acceptance rates drops during the initial burn in phase and finally stabilizes at about \(84\) per cent.
The left panel demonstrates the scatter in the execution times of individual samples. The average execution time is about 300 seconds as indicated by the solid blue line.}
\label{fig:accept_rat}
\end{figure*}

\subsection{Hamiltonian mass}
\label{HMC_MASS}
The HMC possesses a large number of tunable parameters contained in the 'mass' matrix \(M\) which have an important effect on the performance of the sampler.
The Hamiltonian mass defines the inertia  of individual parameters when moving through the parameter space.
Consequently, too large masses will result in slow exploration efficiency, while too light masses will result in large numerical errors of the integration scheme leading to high rejection rates.

Generally, if the individual \(\delta^i_l\) would be Gaussian distributed, a good choice for HMC masses is to set them inversely proportional to the variance of that specific \(\delta^i_l\) \citep{TAYLOR2008}.
For non-Gaussian distributions it is reasonable to use some measure of the width of the distribution \citep{TAYLOR2008}. For example, \citet{NEAL1996} proposes to use the curvature at the peak.
A suitable approach to define Hamiltonian masses is to perform an approximate stability analysis of the numerical leapfrog scheme \citep[see e.g.][]{TAYLOR2008,JASCHE2010HADESMETHOD}. 
Following this approach, we will expand the Hamiltonian forces, given in equation (\ref{eq:Hamiltonian_forces}), around a mean signal \(\xi^i_l\), which is assumed to be the mean initial density contrast once the sampler moved beyond the burn-in phase. As described in Appendix \ref{HAMILTONIAN_MASS} approximating the Hamiltonian forces to linear order amounts to approximating the target distribution by a Gaussian distribution.
According to the discussion in Appendix \ref{HAMILTONIAN_MASS}, the Hamiltonian masses should be set as:
\begin{equation}
\label{eqn:HMC_MASS}
M_{mj}= S^{-1}_{mj} -\delta^K_{mj}\, D^1\,\left. \frac{\partial J_j(s)}{\partial s_j}\right|_{s_j=\xi_j}\, ,
\end{equation}
where \(J_j\) is defined in Appendix \ref{Ap:Hamiltonian_force}.
Calculation of the leapfrog scheme requires inversions of \(M\). Considering the high dimensionality of the problem, inverting and storing \(M^{-1}\) is computationally impractical. For this reason, we construct a diagonal 'mass matrix' from equation (\ref{eqn:HMC_MASS}).
We found that choosing the diagonal of \(M\), as given in equation (\ref{eqn:HMC_MASS}), in its Fourier basis yields faster convergence for the sampler than a real space representation since it accounts for the correlation structure of the underlying density field.

\begin{figure*}
\centering{\includegraphics[width=0.4\textwidth,clip=true]{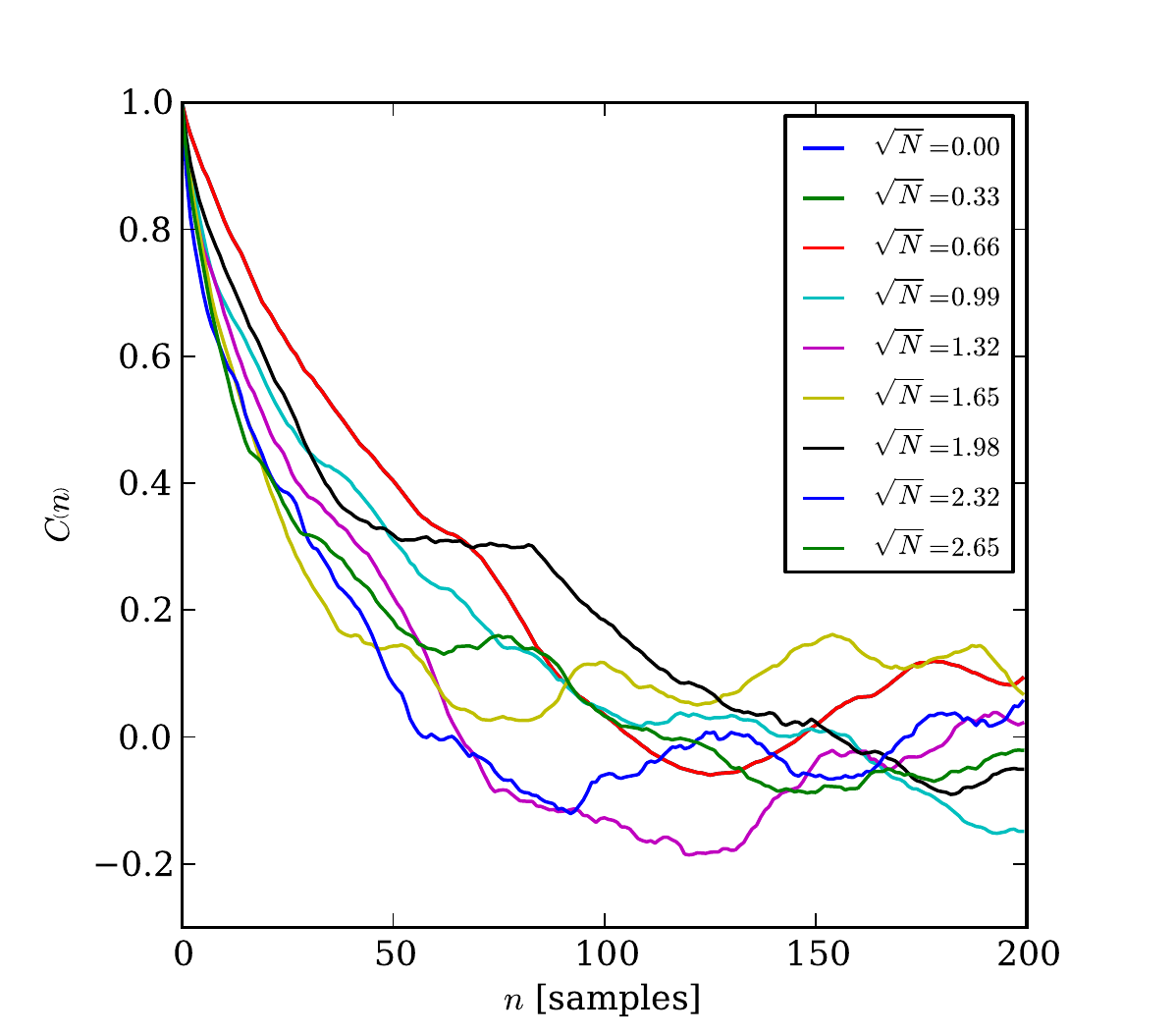}}
\caption{Correlation length for different signal to noise values \(\sqrt{N}\), as indicated in the legend.}
\label{fig:corrlength}
\end{figure*}

\section{Generating Mock observations}
\label{mock_observations}
In the previous sections we presented the derivation and the implementation of our method. Here we will describe the generation of  mock data sets  that will be used to test our method.
Following closely the description in \citet[][]{JASCHE2010HADESMETHOD}, we will first generate a realization for the density contrast \(\delta^i_l\) from a normal distribution with zero mean and a covariance matrix corresponding to a cosmological power-spectrum
in a a three dimensional Cartesian box with \(N_{side}=128\), corresponding to \(N_{vox}=2097152\) volume elements, and a co-moving box length of \(L=750 \mathrm{Mpc}\, h^{-1}\). 
The density field will then be scaled back to an initial time corresponding to a cosmological scale factor \(a_{init}=0.001\) by multiplication with a cosmological growth factor \(D^{+}(a_{init})\), described in appendix \ref{lingrowth}.
In particular, we use a cosmological power-spectrum for the underlying matter distribution, with baryonic wiggles, calculated according to the prescription described in \citet{1998ApJ...496..605E} and \citet{1999ApJ...511....5E}.
We  assume a standard \(\Lambda\)CDM cosmology with a set of cosmological parameters (\(\Omega_m=0.22\), \(\Omega_{\Lambda}=0.78\), \(\Omega_{b}=0.04\), \(h=0.702\), \(\sigma_8=0.807 \), \(n_s=0.961\) ).
The Gaussian initial conditions are then propagated forward in time using second order Lagrangian perturbation theory as described in appendix \ref{2lptmodel}. From the resultant particle distribution
the final density contrast field \(\delta^f_l\) is constructed via the cloud in cell (CIC) method \citep[see e.g.][]{HOCKNEYEASTWOOD1988}.

An artificial galaxy catalog is then generated by simulating the inhomogeneous Poisson process given by equation (\ref{eq:Poissonian}) on top of the final density field \(\delta^f_l\).
In order to set up a realistic testing environment, we seek to emulate the main features of the Sloan Digital Sky survey as closely as possible. We employ the survey geometry of the Sloan Digital Sky Survey
data release 7 depicted in the right panel of figure \ref{fig:TEST_SEL_WIN}. The mask has been calculated using the MANGLE code provided by \citet{SWANSON2008MNRAS} and has been stored on a HEALPIX map with \(n_{side}=4096\) \citep[][]{HEALPIX}. Further, we assume that the radial selection function follows from a standard Schechter luminosity function with standard r-band parameters ( \(\alpha = -1.05\), \(M_* -5 \mathrm{log}_{10}(h)=-20.44\) ), and we only include
galaxies within an apparent Petrosian r-band magnitude range  \(14.5\,<\,r<\,17.77\) and within the absolute magnitude ranges \(M_{min}=-17\) to \(M_{max}=-23\).
The corresponding radial selection function \(f(z)\) is then obtained by integrating the Schechter luminosity function over the range in absolute magnitude:
\begin{equation}
f(z)=\frac{\int^{M_{max}(z)}_{M_{min}(z)} \Phi(M) \, \mathrm{d}M}{ \int^{M_{max}}_{M_{min}} \Phi(M) \, \mathrm{d}M}\, ,
\end{equation}
where \(\Phi(M)\) is given in appendix \ref{schechter_function}.
The resulting selection function for the simulated galaxy sample is depicted in the left panel of figure \ref{fig:TEST_SEL_WIN}.
The survey response operator \(R_i\), required to simulate the Poisson process, can be calculated as the product of the survey geometry and the selection function at each point in the three dimensional volume:
\begin{equation}
R_i = M_i\, F_i= M(\alpha_i,\delta_i) f^l(z_i) \, ,
\end{equation}
Finally, we choose \(\bar{N}=9.93\), and perform the Poisson sampling on the grid resulting in a total number of galaxies \(N_{tot}=484227\).

\begin{figure*}
\centering{\includegraphics[width=1.0\textwidth,clip=true]{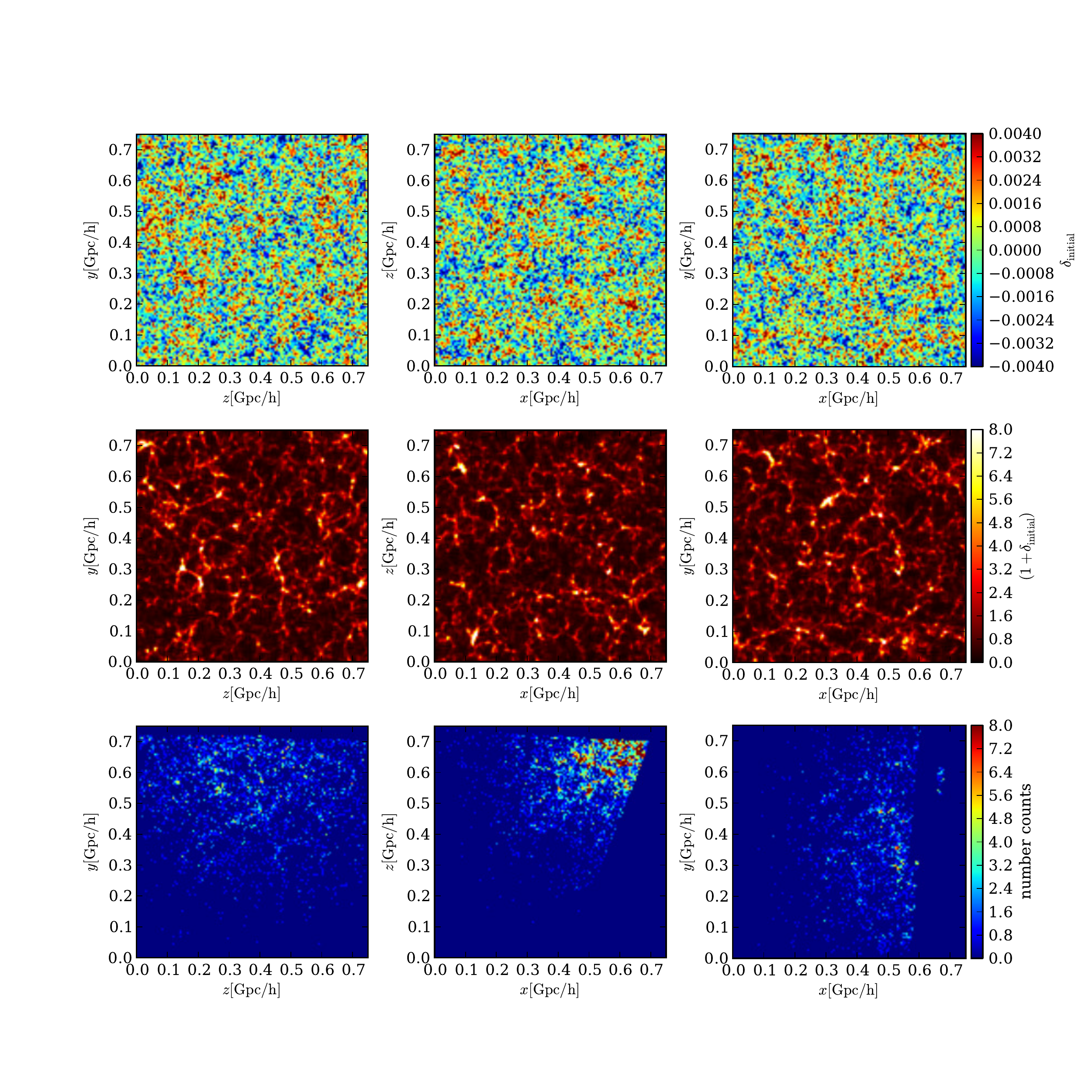}}
\caption{Three slices through a sample of the initial density field (top panels), the final density field (middle panels) and through the corresponding data cube represented by the galaxy number counts (lower panels).
The plots nicely demonstrate the correlation between the final density field and the data. To some extent, one can observe the connection between large structures in the initial conditions and the final density field.}
\label{fig:slices_dens}
\end{figure*}

\section{Testing}
\label{testing}
In this section, we describe the application of our method to the artificial data set described in section \ref{mock_observations}. The primary intention of the following tests is to evaluate the performance of our method in realistic situations, taking into account observational systematics and uncertainties.  

\subsection{Testing convergence and correlations}
\label{convergence}
The Metropolis Hastings Sampler in general and the HMC in particular are designed to have the target distribution, in our case the large scale structure posterior distribution, as its stationary distribution \citep[see e.g.][]{metroplis,hastings,NEAL1993}.
For this reason, the sampling process will provide us with samples from the specified large scale structure posterior distribution after an initial burn-in phase. The length of this initial burn-in phase  has to be assessed using numerical experiments.

Generally, burn-in manifests itself as a systematic drift of the sampled parameters towards the true parameters from which the artificial data set was generated. This behavior can be monitored by following the evolution of parameters in subsequent samples \citep[see e.g.][]{2004ApJS..155..227E,JASCHE2010PSPEC}. 
To test this initial burn-in behavior we will arbitrarily reduce the power of the random initial density field \(\delta^i_l\) by a factor of \(0.1\), and observe the drift towards the true underlying values by following successive power-spectra measured from the samples.
In figure \ref{fig:burn_in} successive power-spectra of the first 800 samples are presented. The plot nicely demonstrates the systematic drift towards the true underlying solution by successive restoration of the true power in the initial density field.

More specifically, we can quantify the initial burn-in behavior by comparing the \(i\)th power-spectrum measurement \(P_i(k)\) in the chain to its true underlying value \(P^{0}(k)\) via:
\begin{equation}
\label{eq:BURN_IN}
\xi\left(P_i(k)\right) = 1.-\frac{P_i(k)}{P^{0}(k)}\, .
\end{equation}
The results for this exercise are presented in the lower panel of figure \ref{fig:burn_in}.
It can be nicely seen that the algorithm restores the correct power an all scales and drifts towards  preferred regions in parameter space to commence exploration of the large scale structure posterior.
It is also important to remark that  in this test, we do not observe any particular hysteresis for the poorly constrained large scale modes, meaning they do not remain at their initially set values but efficiently explore the parameter space. 
This already indicates the ability of our algorithm to account for artificial mode coupling as introduced by the survey geometry.

\begin{figure*}
\centering{\includegraphics[width=1.0\textwidth,clip=true]{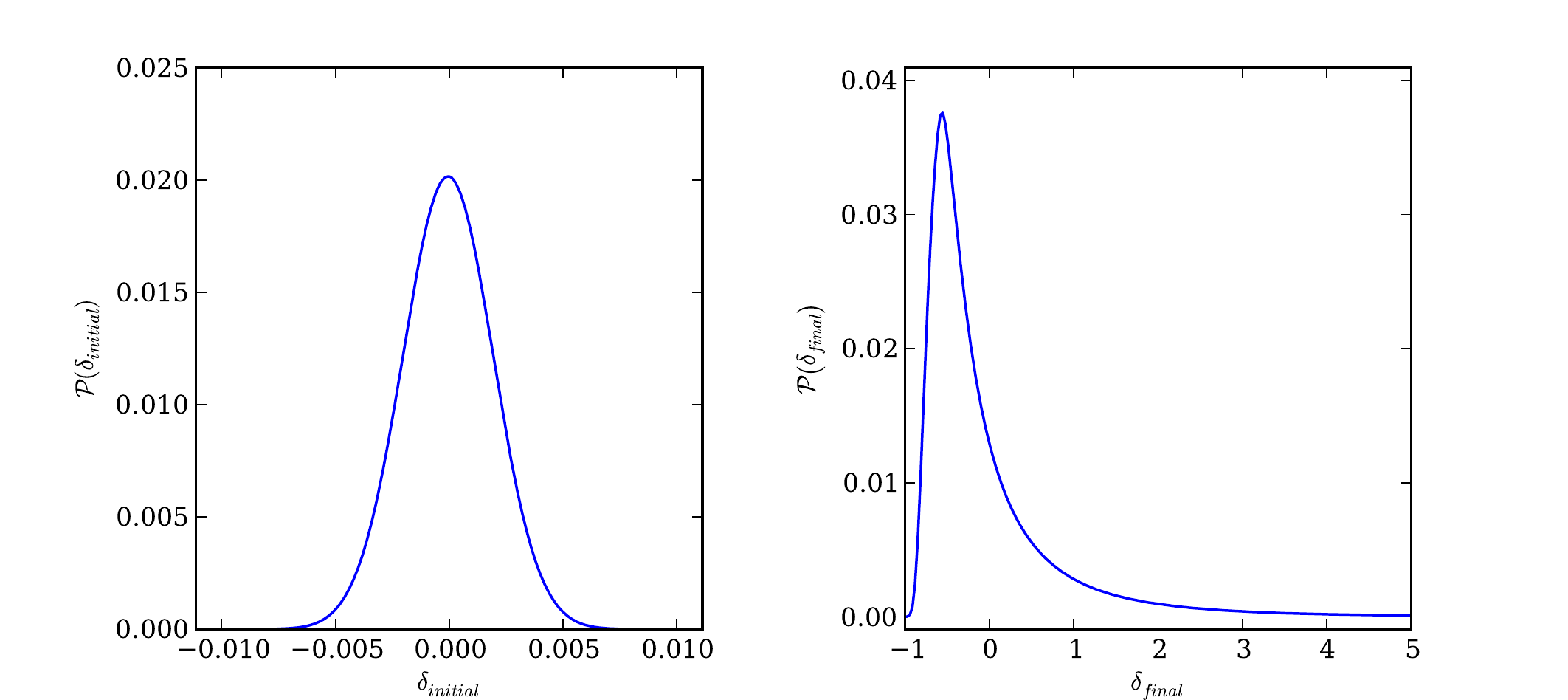}}
\caption{One-point distributions for the density contrast in the initial field (left panel) and for the final field (right panel) measured from the samples. It can be seen that, while the inferred initial density field follows a Gaussian distribution, the final field exhibits the highly skewed log-normal like behavior.}
\label{fig:1pt_pdf}
\end{figure*}

Burn-in also manifests itself in the initial acceptance rate as shown in the left panel of figure \ref{fig:accept_rat}. The dip in the initial acceptance rate function corresponds to the point
when the sampling algorithm restored the full power-of the initial density field. This has a simple explanation. Initially, since the power was reduced by a factor of ten,
the system behaved more or less linear since the displacement of 2LPT particles is small. Once the correct power is restored the displacement of particles increases, leading to a higher non-locality of the system.
When the dip in the acceptance rate occurs, the sampler starts exploring physically more reasonable states in the initial conditions which can explain the  observations. This process is
accompanied by an initially lower acceptance rate. Once the sampler moves to a reasonable region in parameter space the acceptance rate approaches asymptotically a rate of about 84 percent. This high acceptance rate, combined with the fast de-correlation properties, we will demonstrated next, shows that our sampler makes sampling from this multi-million dimensional, non-linear posterior  numerically feasible.

In particular, these tests show that the algorithm requires an initial burn-in phase of on the order of  \(600\) samples before providing samples from the target distribution.
Also note that  the initial burn-in period can be shortened by initializing the algorithm with an initial density field which is closer to the true solution.

Another important issue to study when dealing with Markov Chain Monte Carlo methods is the mixing efficiency of the algorithm.
Generally, successive samples in the chain will not be independent but correlated with previous samples. Consequently, the correlation between successive samples
determines the amount of independent samples which can be drawn from the chain.
We study this effect by following a similar approach as described in \citep[][]{2004ApJS..155..227E} or \citep[][]{JASCHE2010PSPEC}.

Assuming all parameters in the the Markov chain to be independent of one another one can estimate the correlation between subsequent density samples by calculating the autocorrelation function:
\begin{equation}
\label{eq:CORR_COEFF}
C(\delta)_n =\left \langle  \frac{\delta^i-\left \langle \delta\right \rangle}{\sqrt{\mathrm{Var} \left(\delta\right)}} \frac{\delta^{i+n}-\left \langle \delta\right \rangle}{\sqrt{\mathrm{Var} \left(\delta\right)}} \right \rangle \, ,
\end{equation}
where \(n\) is the distance in the chain measured in iterations \citep[also see e.g.][ for a similar discussion]{2004ApJS..155..227E,JASCHE2010PSPEC}.
The results for this analysis are presented in figure \ref{fig:corrlength}, where we plot the correlation coefficients for individual density amplitudes
selected by their signal to noise ratio. It can be generally seen that  the mixing efficiency is a little lower in regions with low signal-to-noise  but the  mixing efficiency of the algorithm
is very good overall.

We can further define a correlation length of the Markov sampler as the distance in the chain \(n_c\) beyond which the correlation coefficient \(C(\delta)_n\) has dropped below a threshold of \(C^{th}(\delta)_n=0.1\). 
As can be seen in figure \ref{fig:corrlength} the correlation length is about 200 samples, demonstrating the high mixing efficiency of the algorithm.
Despite the high dimensionality of the problem considered here, these tests demonstrate that exploring large scale structure posterior for the initial conditions
from observations exhibiting systematic uncertainties and uncertainties is numerically feasible with our method.

\subsection{Large Scale Structure inference}
\label{lss_inference}
In this section we will discuss the results obtained from the application of the algorithm to the artificial data set, as described in section \ref{mock_observations}.
The initial and final density fields have been inferred on a \(128^3\) cubic Cartesian box with side length of \(750 \, \mathrm{Mpc}/h\), resulting in a grid resolution of 
about \(\sim 6 \mathrm{Mpc}/h\).  
While sampling the algorithm will provide matter field realizations for the initial and final 2LPT density fields, generated conditionally on the observed data.

In figure \ref{fig:slices_dens} we show slices from three different sides through samples of the initial and corresponding final densities as well as through the data.
It is immediately visible that the statistics of the initial and final matter fields  differ greatly. While the initial density field appears to be very Gaussian,
the final density field is clearly non-Gaussian. This demonstrates how the physical 2LPT model for structure formation is able to account for 
the growing statistical complexity of the density distribution during the evolution from the initial to the final state.
Furthermore, comparison of the final density field (middle panels of figure \ref{fig:slices_dens}) to the data (lower panels of figure \ref{fig:slices_dens})
demonstrates the recovered structures from the data. One can nicely see that the algorithm tries to extrapolate unobserved filaments between clusters
based on the physically reasonable assumptions provided by the 2LPT model. In general, the algorithm nicely recovers the filamentary structure of the matter distribution.

Figure \ref{fig:slices_dens} illustrates that the algorithm accurately accounts for survey geometry and selection effects
by augmenting unobserved or poorly observed regions with statistically correct information. The inferred initial and final
density fields possess equal power throughout their entire domains and are not affected by selection or mask artifacts. Individual density samples can be understood
as  physical, three-dimensional matter field realizations, at least to the degree permitted by the 2LPT model.
It is particularly interesting that unobserved and observed regions in the inferred final density fields do not appear visually distinct, a consequence of the excellent approximation of the 2LPT not just to the first but also  higher order moments  \citep[][]{MOUTARDE1991,BUCHERT1994,BOUCHET1995,SCOCCIMARRO2000,PTHALOS}.
This is a great advantage over previous methods based on Gaussian or log-normal models where the assumption of a cosmological power-spectrum only specifies the two-point statistics correctly. In particular, the reader may want to compare with figure 2 in \citet{JASCHE2010HADESDATA}, where unobserved regions are augmented with a log-normal model unable
to represent filamentary structures.

\begin{figure*}
\centering{\includegraphics[width=0.8\textwidth,clip=true]{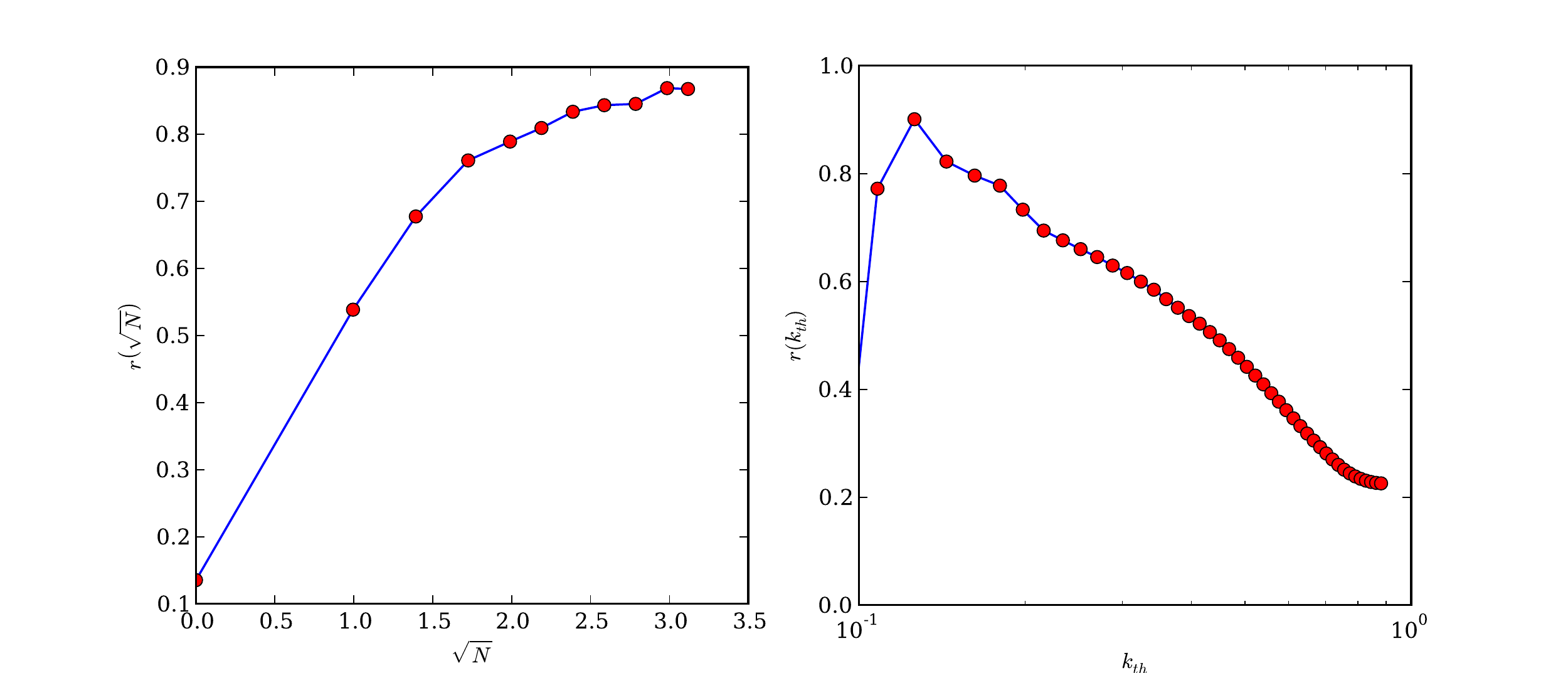}}
\caption{Cross correlation coefficient between the true final density field and a sample as a function of signal to noise (left panel) and the cross corelation between the true underlying initial density field and the inferred ensemble mean initial field as a function of smoothing scale \(k_{th}\) (right panel). It is interesting to remark, that the correlation between true underlying and samples of the density field still amounts to about 55 per cent in regions where only a single galaxy has been observed.}
\label{fig:cross_corr_rec2truth}
\end{figure*}

In figure \ref{fig:1pt_pdf} we compare the one-point distribution of the inferred initial and final density field measured from the corresponding samples. It can be seen that while the initial density contrast follows Gaussian statistics, the final distribution
is highly skewed and represents the expected log-normal like behavior. These results therefore supports our initial claim that  the complex
problem of modeling a prior distribution for the present fully non-linear density field can be exchanged for an initial conditions inference problem when assuming a physical model
which accounts for the increasing statistical complexity of the matter distribution during structure formation.

Further, we estimate the accuracy of the recovered density field by estimating the correlation coefficient \(r(x)\) between density samples and the true underlying solution as a function of some parameter \(x\).
The correlation coefficient is given as: 
\begin{equation}
r(k_{x})=\frac{\left\langle \delta_{0}^{x} \, \langle\delta \rangle^{x} \right \rangle}{ \sqrt{\langle \left(\delta_{0}^{x}\right)^2 \rangle}\,\sqrt{\langle \left(\langle\delta \rangle^{x}\right)^2 \rangle}}\, ,
\end{equation}
where we will choose \(x\) to be the signal to noise ratio \(\sqrt{N}\) for the final density field and a specific smoothing scale \(k_{th}\) for the initial density field.
The results for these tests are demonstrated in figure \ref{fig:cross_corr_rec2truth}.  The left panel of figure \ref{fig:cross_corr_rec2truth} depicts the correlation
between the true underlying final density field and the final density samples as a function of the signal to noise ratio.
It can be seen that  the correlation with the truth is generally higher for higher signal to noise ratios. Even in zones that contain just a single galaxy we still
get a correlation of about 55 percent. It is also remarkable that  the algorithm still provides a 10 percent correlation
with the true underlying density field in regions which have not been sampled by galaxies such as centers of voids or masked regions.
The right panel of figure \ref{fig:cross_corr_rec2truth} demonstrates the cross correlation between the true underlying initial conditions and the inferred ensemble
mean initial field as a function of filter scale \(k_{th}\) when smoothed with a spherical top hat filter in Fourier space.
These results clearly demonstrate that the large scales of the initial conditions can be much easier recovered than the small scale features.
This is in agreement with the expectation, since the largest scales behave more linearly than the smaller scales and hence are easier to recover.
Particular the shot noise contribution at the smallest scales in the final galaxy observation will smear out features in the initial conditions,
since the 2LPT displacement vector for the particles will fluctuate on these scales.

\subsubsection{Dynamics} 
\begin{figure*}
\centering{\includegraphics[width=1.1\textwidth,clip=true]{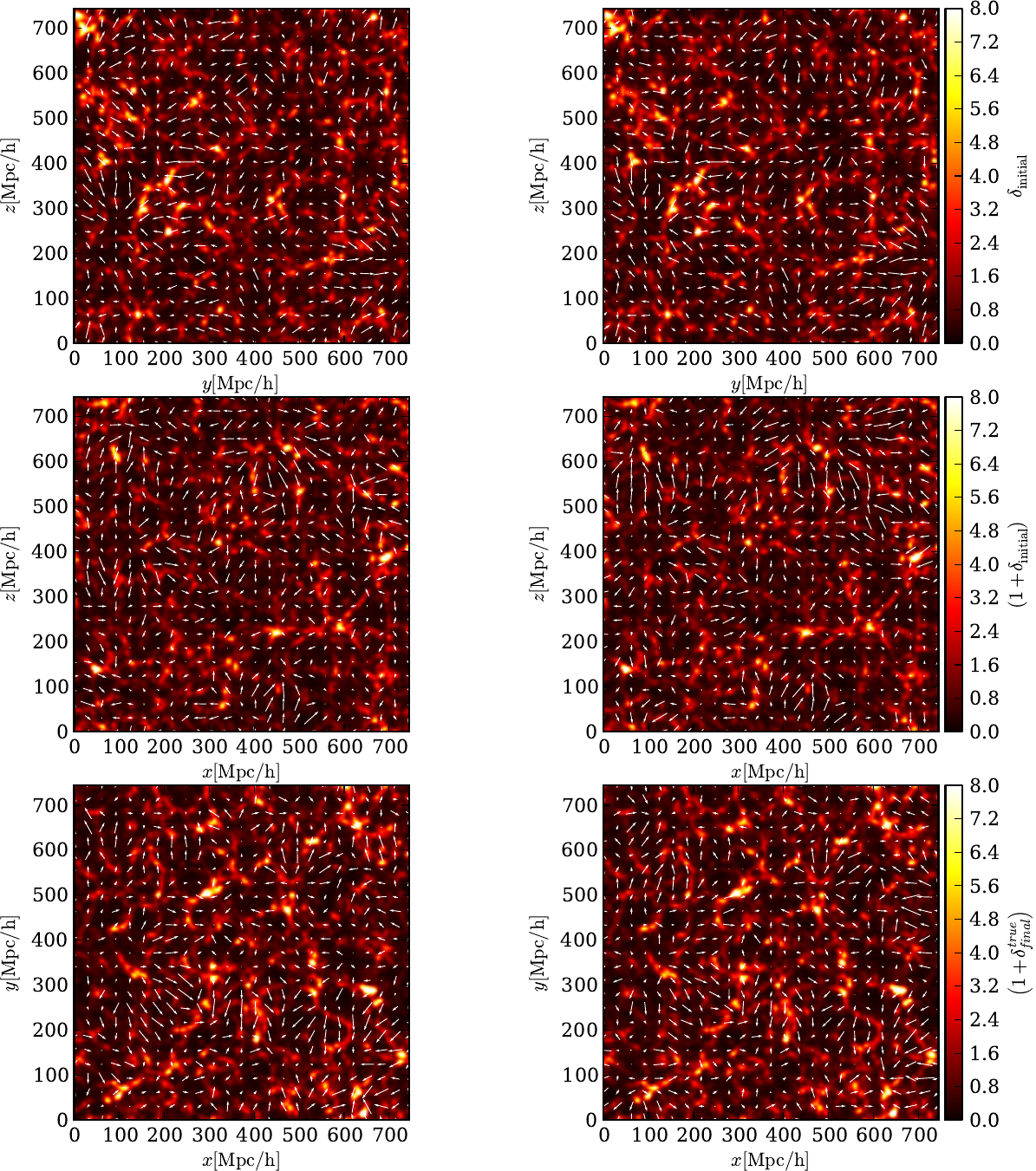}}
\caption{Three slices through the true underlying density field from three different sides over plotted with the two dimensional projection of the true velocity (left panels) and a sample velocity field (right panels). It can be seen that the algorithm is able to infer the true underlying dynamics of the system to great detail in noisy and even unobserved regions, when compared to the corresponding data panels in figure \ref{fig:slices_dens}.}
\label{fig:slices_dens_velocity}
\end{figure*}

Importantly, the algorithm provides dynamical information on the large scale structure given the 2LPT model.
In figure \ref{fig:slices_dens_velocity}, we show the comparison between the true underlying velocity field, and the velocity field
inferred by an randomly selected sample. It can be seen that  the algorithm is able to recover the true underlying velocity field in detail.
Our inference is clearly able to identify the true underlying velocity field in noisy or even completely masked regions.
This clearly demonstrates the strength of this approach in extrapolating physically reasonable states of the matter distribution even into poorly observed regions.

\section{Discussion and Conclusion}
\label{Conclusion}
We describe a new method to perform dynamical large scale structure inference from galaxy redshift surveys employing a second order Lagrangian
perturbation model. 
In section \ref{2lpt_posterior} we demonstrated that the problem of constructing suitable prior distributions for the non-linear density field is directly linked
to the problem of inferring initial conditions, once a dynamical model for large scale structure formation is given. 
In this approach the evolved non-linear density field acts as a mere nuisance parameter in the inference process, which shifts the problem of designing
prior distributions to physical modeling of the matter evolution dynamics.

Since the method we propose provides an approximation to the non-linear dynamics the algorithm automatically provides information on the dynamical evolution of the large scale matter distribution. 
By exploring the space of   dynamical \textit{histories} compatible with both  data and  model our approach therefore marks the passage from Bayesian three-dimensional density inference to full four-dimensional state inference.

Particularly, in this work we have employed a 2LPT model as an approximate dynamical description of the large scale structure evolution on the large scales.
As described in the literature, the 2LPT model describes the one, two and three-point statistics correctly and represents higher order statistics very well \citep[see e.g. ][]{MOUTARDE1991,BUCHERT1994,BOUCHET1995,SCOCCIMARRO2000,PTHALOS}.
Hence, the algorithm proposed in this work can exploit higher order statistics, modeled through the 2LPT model, to provide physically reasonable matter field
realizations conditional on the observed galaxy distribution.

It is also important to remark that  the inference process described in section \ref{2lpt_posterior} requires at no point the inversion of the flow of time
in the dynamical model. The inference process therefore solely depends on forward propagation of the model, which consequently alleviates many of the problems
encountered in previous approaches to the reconstruction of initial conditions, such as spurious decaying mode amplification.
Rather than inferring  the initial conditions by backward integration in time our approach builds a non-linear filter using the dynamical forward model as a prior. This prior  singles out physically reasonable large scale structure states
from the space of all possible solutions.

The resultant inference procedure is numerically highly non-trivial, since the large scale structure posterior distribution has to be evaluated in very high dimensional space.
Typically we are dealing with \(10^6\) to \(10^7\) parameters, corresponding to the voxels used to discretize the domain. 
In section \ref{HAMILTONIAN_SAMPLING}, we described an efficient Hybrid Monte Carlo implementation for the large scale structure inference problem when employing a dynamical model for large scale structure formation.
Further, we discussed some details of the numerical implementation in section \ref{Numerical_implementation}.

To provide a proof of concept we test the algorithm in an artificial  scenario, based on the characteristics of the Sloan Digital Sky Survey Data Release 7.
In particular, as described in section \ref{mock_observations}, we use the SDSS DR7 completeness map and realistic selection functions based on the Schechter luminosity function
to generate a realistic testing environment essentially emulating the SDSS DR7 main sample.  

The major aim of testing the algorithm, described in section \ref{testing}, was  to estimate the method's performance in a realistic scenario.
An important issue to test when dealing with Markov Chain Monte Carlo methods is the question of how many independent samples can be drawn from the chain.
The high efficiency of our  Hybrid Monte Carlo scheme permits to explore the posterior distribution with a typical acceptance rate of about \(84\) per cent while maintaining the correlation length of the chain at or below $300$ steps. We estimate the length of the burn-in phase to be about 600 steps. 
In summary, our tests reveal that the proposed analysis approach is not only within numerical reach but is efficient enough to work well with present day computational resources.

The properties of the inferred large scale structure fields were studied in section \ref{lss_inference}.
It is clear upon visual inspection that our approach returns far more physical reconstructions than previous methods based solely on two-point information \citep[see e.g.][]{LAHAV1994,ZAROUBI2002,ERDOGDU2004,KITAURA2008MNRAS,KITAURA2009MNRAS,KITAURA2010MNRAS,JASCHE2010PSPEC,JASCHE2010HADESMETHOD,JASCHE2010HADESDATA,JASCHE2010PSPEC}.
This is particularly obvious for unobserved regions which are augmented with statistically correct information, in order to account for survey geometry and cosmic variance.
In the present approach augmented regions are visually indistinguishable from regions containing data. Therefore, the individual matter field samples can be regarded as full physical matter field realizations conditional on the observations, at least to the degree represented by the 2LPT model.

By studying the one-point distributions of the inferred initial and final density fields we demonstrated that the algorithm correctly recovers the Gaussian initial
conditions from a galaxy observation which does not exhibit Gaussian but highly skewed log normal like statistics. This demonstrates that the algorithm correctly,
accounts for the mode coupling and phase correlations originally introduced to the matter distribution by gravitational structure formation.
In addition, it supports our initial claim that  the approach of searching for phenomenological approximations to the full probability distribution for the non-linear
matter field can be efficiently reformulated as an initial condition problem once a physical model for large scale structure is employed.

To estimate the accuracy of recovered density fields, we studied the correlation between the true underlying and samples of the final density field as a function of the signal to noise ratios.
As expected, the correlation can reach up to 90 per cent in the high signal to noise regime, where $S/N\sim3$. In addition, the algorithm
still provides a correlation of about 50 per cent between the true underlying density field and the samples  in regions where only a single galaxy has been observed.
Also note that in regions where the signal to noise is zero, which are either centers of voids or unobserved regions, the algorithm still provides a 10 per cent 
correlation. This is a clear manifestation of improved inference due to the incorporation of a physical model of large scale structure formation, which exploits additionally three point
and higher moment statistics of the density distribution.   
These tests further demonstrate that the algorithm correctly accounts for systematics such as the survey geometry and selection effects.

Along with the inferred density fields the algorithm also provides dynamical information on the large scale flows. By comparing the true underlying velocity field
to the inferred velocity field of an arbitrary sample we demonstrated that  the algorithm accurately recovers large scale flows, even in noisy or even unobserved regions.
This clearly demonstrates the strength of the method in extrapolating physically reasonable states into poorly observed regions.

The method we describe forms the basis for a sophisticated and extensible dynamical large scale structure inference framework. 
In  future work we will demonstrate the application of the algorithm to a real galaxy survey accounting for additional systematics 
such as luminosity or color dependent bias. Note that the algorithm as described in this work can be easily extended to account for any kind of non-linear
and non-local bias. In particular, the 2LPT model, as employed in this work, can already be interpreted as a non-local, non-linear bias model between the initial
conditions and the galaxy observations. It would also be possible to incorporate a halo model based galaxy bias model in the fashion as described by \citet{PTHALOS}.
The combination of the algorithm described in this work and the photometric redshift sampling method proposed in \cite{JASCHE2012}, will lead to immediate improvements for the inferred photometric redshifts,
since the combination of both algorithms will exploit higher order statistics, whereas the algorithm described in \cite{JASCHE2012} is solely based on two-point statistics.
In a similar fashion, dynamical velocity information provided by the 2LPT model can be used to correct for redshift uncertainties in spectroscopic surveys.

Since the algorithm is fully Bayesian, it provides inferred initial and final density fields and also the means of estimating their significance and uncertainties by
a sampled representation of the initial conditions posterior distribution. The algorithm will therefore provide accurate information on the initial conditions from which the observed large scale structure originates.
These initial density fields may be useful for a variety of scientific projects such as constrained simulations \citep[see e.g.][]{KRAVTSOV2002,KLYPIN2003,DOLAG2005,LIBESKIND2010,GOTTLOEBER2009,GOTTLOEBER2010,LAVAUX2010,DOLAG2012}.
Since the 2LPT model reconstructs the initial tidal field it may also open up a new way to study the angular momentum build-up of galaxies through tidal torque theory \citep[see e.g. the review by ][ and references therein]{SCHAEFER2009}.

In conclusion, we presented a new Bayesian dynamical large scale structure inference algorithm which will provide the community with
accurate measurements of the three dimensional initial density field as well as estimates of the dynamical behavior of the large scale structure.

\section*{Acknowledgments}
JJ is partially supported by a Feodor Lynen Fellowship by the Alexander von Humboldt foundation. BDW acknowledges  support from NSF grants AST 07-08849 and AST 09-08693 ARRA, and a Chaire d'Excellence from the Agence Nationale de Recherche and computational resources provided through  XSEDE grant AST100029.

\bibliography{paper}
\bibliographystyle{mn2e}

\appendix
\section[]{Linear structure formation}
\label{lingrowth}
In the linear regime structure formation is governed by a homogeneous growth function \(D^{+}(a)\) acting on the density contrast \(\delta(\vec{x},a)=D^{+}(a)\,\delta(\vec{x},a=1)\).
For a general cosmology the growth factor \(D^{+}(a)\) can be obtained by numerical solution of the linear growth equation \citep[see e.g.][]{TURNER1997,WANG1998,LINDER2003}:
\begin{equation}
\label{eq:growth_factor}
\frac{\mathrm{d}^2D^{+}(a)}{\mathrm{d}a^2}+ \frac{1}{a}\left(3+\frac{\mathrm{d\,ln} H}{\mathrm{d\,ln} a}\right)\frac{\mathrm{d}D^{+}(a)}{\mathrm{d}a}-\frac{3}{2}\frac{\Omega_m(a)D^{+}(a)}{a^2}=0
\end{equation}

\section[]{Lagrangian Perturbation theory}
\label{2lptmodel}
In the following we will give a brief summary of second order Lagrangian perturbation theory to the degree required for the present work. More detailed discussion of Lagrangian perturbation theory in general and its application can be found in the literature \citep[see e.g.][]{MOUTARDE1991,BUCHERT1994,BOUCHET1995,SCOCCIMARRO1998,SCOCCIMARRO2000,PTHALOS,BERNADEAU2002}.
Also see \citet{BERNADEAU2002} for a general overview of Eulerian and Lagrangian cosmological perturbation theory.

In an expanding Robertson Friedman space time the equations of motion for particles solely interacting through gravity are given as \citep[see e.g. ][]{BERNADEAU2002,SCOCCIMARRO2000}:
\begin{equation}
\label{eq:equation_of_motion}
\frac{\mathrm{d}^2\vec{x}}{\mathrm{d}\tau^2}+  \mathcal{H} \frac{\mathrm{d}\vec{x}}{\mathrm{d}\tau}-\nabla_x \phi=0 \, ,
\end{equation}
where \(\Phi\) is the gravitational potential and \(\nabla_x\) is the gradient with respect to the Eulerian coordinates \(\vec{x}\), \(\mathcal{H}=\mathrm{d\,ln\,a}/\mathrm{d\tau}\) and the conformal time \(\tau\) defined by \(\mathrm{d\tau}= \mathrm{dt}/a\) .
In order to solve this set of equations, Lagrangian perturbation theory introduces the following Ansatz for a solution:
\begin{equation}
\label{eq:equation_of_motion_ansatz}
\vec{x}(\tau)=\vec{q}+\vec{\Psi}(\vec{q},\tau)\, ,
\end{equation}
where \(\vec{\Psi}(\vec{q},\tau)\) defines the mapping from the particles initial position \(\vec{q}\) into its final Eulerian position \(\vec{x}\) \citep[see e.g. ][]{BERNADEAU2002,SCOCCIMARRO2000}. Equation (\ref{eq:equation_of_motion_ansatz}) 
together with equation (\ref{eq:equation_of_motion}) yields a non-linear equation for the displacement field \(\vec{\Psi}(\vec{q},\tau)\) which can be solved perturbatively by expanding around its linear solution 
\citep[][]{BERNADEAU2002}.
To linear order, this perturbative approach yields the famous Zel'dovich approximation given as \citep[][]{ZELDOVICH1970,DOROSHKEVICH1970,BUCHERT1989,MOUTARDE1991,BERNADEAU2002}:
\begin{equation}
\label{eq:disp_1lpt}
\nabla_q \Psi^{(1)}(\vec{q},a) = - D^{+}(a)\, \delta\left(\vec{q}, a=1\right)\, .
\end{equation}
Adding second order terms to the perturbative expansion remarkably improves the results of the first order Zel'dovich approximation. In particular, second order
terms account for the fact that gravitational instability is non-local by introducing corrections due to gravitational tidal effects \citep[][]{BERNADEAU2002}. The second order displacement
field \(\Psi^{(2)}(\vec{q},a)\) is then defined by \citep[see e.g. ][]{BERNADEAU2002,PTHALOS}:
\begin{equation}
\label{eq:disp_2lpt}
\nabla_q \Psi^{(2)}(\vec{q},a) = \frac{1}{2} D_2(a)\, \sum_{i\neq j} \left( \Psi^{(1)}_{i,i}\, \Psi^{(1)}_{j,j} - \Psi^{(1)}_{i,j}\,\Psi^{(1)}_{j,i}\right)\, ,
\end{equation}
with \(\Psi^{(1)}_{i,j} \equiv \partial \Psi^{(1)}_{i} / \partial \vec{q}_j \) and \(D_2(a)\) is the second order growth factor given as:
\begin{equation}
\label{eq:growthfactor_2lpt}
D_2(a) \approx -\frac{3}{7} \left(D^{+}(a)\right)^2 \, \Omega_m^{-\frac{1}{143}}\, ,
\end{equation}
which holds for a flat model with non-zero cosmological constant \(\Lambda\) and for \(0.01\leq\Omega_m \leq1\) to better than \(0.6\) per cent accuracy \citep[see e.g. ][ for details]{BOUCHET1995,SCOCCIMARRO1998,BERNADEAU2002}.

As has been previously shown, second order Lagrangian perturbation theory recovers correctly the two- and three-point statistics at large scales and further approximates higher-order statistics very well
\citep[][]{MOUTARDE1991,BUCHERT1994,BOUCHET1995,SCOCCIMARRO2000,PTHALOS}. 
Also note, that second order corrections to the Zel'dovich approximation are essential to accurately describe the departure of the large scale density field from Gaussian initial conditions \citep[][]{PTHALOS,TATEKAMA2007,JENKINS2010}.

Lagrangian solutions up to second order are curl free, as they follow potential flows \citep[see e.g. ][]{BUCHERT1994,SCOCCIMARRO1998,BERNADEAU2002}. Therefore, it is convenient to introduce the Lagrangian potentials \(\Phi^{(1)}\) and \(\Phi^{(2)}\),
such that the approximate solution to equation (\ref{eq:equation_of_motion}) can be expressed as \citep[see e.g. ][]{BUCHERT1994,SCOCCIMARRO1998,BERNADEAU2002}: 
\begin{equation}
\label{eq:x_2lpt}
\vec{x}(\tau)=\vec{q} - D^{+}(a) \nabla_q\, \Phi^{(1)}+ D_2 \Phi^{(2)}\, ,
\end{equation}
where the time-independent potentials \(\Phi^{(1)}\) and \(\Phi^{(2)}\) are solutions to the following Poisson equations \citep[][]{BUCHERT1994}:
\begin{equation}
\label{eq:Poisson_1lpt}
\nabla^2_q\, \Phi^{(1)}(\vec{q})=\delta\left(\vec{q}, a=1\right)\, ,
\end{equation}
and
\begin{equation}
\label{eq:Poisson_2lpt}
\nabla^2_q\, \Phi^{(2)}(\vec{q})=\sum_{i>j} \left[ \Phi^{(1)}_{,ii}(\vec{q})\,\Phi^{(1)}_{,jj}(\vec{q}) - \left( \Phi^{(1)}_{,ij}(\vec{q}) \right)^2 \right]\, ,
\end{equation}
For an excellent guide to the numerical implementation of the 2LPT model the reader is referred to appendix D of \citet{SCOCCIMARRO1998}.

\section{The Schechter Luminosity function}
\label{schechter_function}
The Schechter luminosity function is given as \citep[][]{SCHECHTER1976}:
\begin{equation}
\Phi(M)\, \mathrm{d}M = 0.4\, \Phi^{*} \mathrm{ln}(10)\, \left(10^{0.4\,(M^{*}-M)}\right)^{\alpha+1}\, \mathrm{e}^{-10^{0.4\,(M^{*}-M)}}\,\mathrm{d}M \, . 
\end{equation}
Note that for the purpose of calculating selection functions the normalization \(\Phi^{*}\) is not required.

\section{Hamiltonian Forces for the likelihood term}
\label{Ap:Hamiltonian_force}
In this section we will discuss the derivation of the Hamiltonian forces for the 2lpt-Poissonian process.
To prevent confusion between the variables describing the physical 2LPT model and the variables describing the Hamiltonian inference framework
we will re express the 2LPT model in the following form for the purpose of the derivations in this section:
\begin{equation}
\label{eq:2lpt_position}
\vec{x}_p = \vec{x}_p(\delta^i) = \vec{q}_p - D^1\, \vec{K}^1_p(\delta^i) + D_2\, \vec{K}^2_p(\delta^i) \,  ,
\end{equation}
where \(\vec{K}^1_p(\delta^i)\) and \(\vec{K}^2_p(\delta^i)\) are the first and second order displacements fields, respectively.

As described in section \ref{equations_of_motion} the likelihood term of the Hamiltonian potential is given as:
\begin{eqnarray}
\Psi_{likelihood}(\{\delta^i_j\}) &=& \sum_l R_l\bar{N}_{gal}\,(1+G(a,\delta^i)_l) \nonumber \\
& &- N_l {\rm{ln}}\left(R_l\bar{N}_{gal}\,(1+G(a,\delta^i)_l)\right) \, ,
\end{eqnarray}
with \(G(a,s)\) given via the kernel estimate as:
\begin{equation}
\label{eq:deltai}
G(a,s)_l = \sum_p \frac{W(\vec{x}_p(a,s) -\vec{x}_l)}{\bar{N}} - 1\, ,
\end{equation}
and \(\vec{x}_p(a,s)\) is described by equation \ref{eq:2lpt_position} and \(W(\vec{x})\) is a CIC kernel \citep[see e.g.][]{HOCKNEYEASTWOOD1988,JASCHE2009DSPC}.
Furthermore, the Lagrangian displacement vectors are given as:
\begin{equation}
\label{eq:displacement_vector}
K^n_p = \sum_j \vec{V}'(\vec{q}_p-\vec{x}_j) \Phi^n_j \,  ,
\end{equation}
where \(\vec{V}'(\vec{x})\) is the gradient of the kernel \(W(\vec{x})\).
With these definitions we can write the Hamiltonian forces corresponding to the likelihood term as:
\begin{eqnarray}
\label{eq:LH_Forces}
\frac{\partial\Psi_{likelihood}(\{\delta^i_j\})}{\partial s_m}  &=&\sum_i \left( 1 - \frac{1}{R_i\, \bar{N}\,(1+G(a,\delta^i)_l)} \right) R_i\, \bar{N} \frac{\partial G(a,\delta^i)_i)}{\partial \delta^i_m} \, .\nonumber \\
\end{eqnarray}
The notation can be simplified by introduce the quantity \(A_i\) as:
\begin{equation}
\label{eq:Ai}
A_i =  \left( 1 - \frac{1}{R_i\, \bar{N}\,(1+\delta_i(s)} \right) R_i\, \bar{N} \, .
\end{equation}
We can then write:
\begin{eqnarray}
\label{eq:LH_Forces_a}
\frac{\partial\Psi_{likelihood}(\{\delta^i_j\})}{\partial \delta^i_m} &=&\sum_i A_i \frac{\partial G(a,\delta^i)_i}{\partial \delta^i_m} \nonumber \\
&=&\sum_i \frac{A_i}{\bar{N}} \sum_p \frac{\partial W(\vec{x}_p -\vec{x}_i)}{\partial \delta^i_m} \nonumber \\
&=&\sum_i \frac{A_i}{\bar{N}} \sum_p \vec{W}'(\vec{x}_p -\vec{x}_i) \frac{\partial \vec{x}_p}{\partial \delta^i_m} \nonumber \\
&=&\sum_i \frac{A_i}{\bar{N}} \sum_p \vec{W}'(\vec{x}_p -\vec{x}_i) \left(  - D^1\,\frac{\partial K^1_p(\delta^i)}{\partial \delta^i_m} \right . \nonumber \\
& & \left . +D^2\,\frac{\partial K^2_p(\delta^i)}{\partial \delta^i_m} \right)\, , \nonumber \\
\end{eqnarray}
where we made use of equations (\ref{eq:2lpt_position}) and (\ref{eq:displacement_vector}).
It can be seen that the Hamiltonian force is the sum of two vectors. In the following we will therefore discuss each term independently.
The first term is exactly the Hamiltonian force expected from a pure Zeldovich approximation without higher order correction terms.
We will start by evaluating the Hamiltonian force for the Zeldovich approximation.
\begin{eqnarray}
\label{eq:LH_Forces_b}
\frac{\partial\Psi^1_{likelihood}(\{\delta^i_j\})}{\partial \delta^i_m} &=&\sum_i \frac{ - D^1\,A_i}{\bar{N}} \sum_p \vec{W}'(\vec{x}_p -\vec{x}_i) \left( \,\frac{\partial \vec{K}^1_p(\delta^i)}{\partial \delta^i_m} \right)\, , \nonumber \\
&=& \sum_p \sum_i \frac{ - D^1\,A_i}{\bar{N}}  \vec{W}'(\vec{x}_p -\vec{x}_i)  \sum_j \vec{V}'(\vec{q}_p-\vec{x}_j) \frac{\Phi^1_j}{\partial \delta^i_m}  \nonumber \\
&=& \sum_j \sum_p \sum_i \frac{ - D^1\,A_i}{\bar{N}}  \vec{W}'(\vec{x}_p -\vec{x}_i)   \vec{V}'(\vec{q}_p-\vec{x}_j) \frac{\Phi^1_j}{\partial \delta^i_m} \, ,  \nonumber \\
\end{eqnarray}
The notation can be further simplified by introducing:
\begin{equation}
\label{eq:Fj}
F_j =  \sum_p \sum_i \frac{A_i}{\bar{N}}  \vec{W}'(\vec{x}_p -\vec{x}_i)   \vec{V}'(\vec{q}_p-\vec{x}_j) \, .
\end{equation}
We can then write
\begin{eqnarray}
\label{eq:LH_Forces_c}
\frac{\partial\Psi^1_{likelihood}(\{\delta^i_j\})}{\partial \delta^i_m} &=& - D^1 \sum_j F_j \frac{\Phi^1_j}{\partial \delta^i_m} \, . 
\end{eqnarray}
The Zeldovich Approximation potential was calculated using the Fast Fourier Transform approach, which can be written as:
\begin{eqnarray}
\label{eq:Phi_ZA}
\Phi^1_j &=& \sum_k \frac{-1}{k_k^2} {\rm{e}}^{2\pi j\,k \frac{\sqrt{-1}}{N}}  \sum_n s_n {\rm{e}}^{-2\pi n\,k \frac{\sqrt{-1}}{N}}  \, . 
\end{eqnarray}
Using this expression in equation (\ref{eq:LH_Forces_b}) we yield:
\begin{eqnarray}
\label{eq:LH_Forces_d}
\frac{\partial\Psi^1_{likelihood}(\{\delta^i_j\})}{\partial \delta^i_m} &=& - D^1\sum_j F_j  \sum_k \frac{-1}{k_k^2} {\rm{e}}^{2\pi j\,k \frac{\sqrt{-1}}{N}}  \sum_n \delta^K_{nm} {\rm{e}}^{-2\pi n\,k \frac{\sqrt{-1}}{N}}\, \nonumber \\ 
&=& - D^1\sum_j F_j  \sum_k \frac{-1}{k_k^2} {\rm{e}}^{2\pi j\,k \frac{\sqrt{-1}}{N}} {\rm{e}}^{-2\pi m\,k \frac{\sqrt{-1}}{N}}\, \nonumber \\
&=&- D^1 \sum_k \frac{-1}{k_k^2} {\rm{e}}^{-2\pi m\,k \frac{\sqrt{-1}}{N}} \sum_j F_j\,  {\rm{e}}^{2\pi j\,k \frac{\sqrt{-1}}{N}} \, . \nonumber \\ 
\end{eqnarray}
This result looks remarkably similar to equation (\ref{eq:Phi_ZA}) and at first sight one might be inclined to straightforwardly solve this equation with FFT techniques.
However, it is important to note that the signs have changed in the exponents, and hence equation (\ref{eq:LH_Forces_b}) can not directly be solved with FFTs. In Appendix \ref{USE_FFTS}, we show what procedures must be followed in order to apply FFTs to this problem.
To further simplify the notation in the following steps we will introduce the quantity \(J_m\), defined as:
\begin{eqnarray}
\label{eq:JM}
J_m &=& \sum_k \frac{-1}{k_k^2} {\rm{e}}^{-2\pi m\,k \frac{\sqrt{-1}}{N}} \sum_j F_j\,  {\rm{e}}^{2\pi j\,k \frac{\sqrt{-1}}{N}} \, . 
\end{eqnarray}
With this definition the ZA term of the Hamiltonian force can be written as:
\begin{eqnarray}
\label{eq:LH_Forces_e}
\frac{\partial\Psi^1_{likelihood}(\{\delta^i_j\})}{\partial \delta^i_m} &=& - D^1\, J_m \, . 
\end{eqnarray}

Next, we will discuss the second order Lagrangian term in equation (\ref{eq:LH_Forces_a}):
\begin{eqnarray}
\label{eq:LH_Forces_b2}
\frac{\partial\Psi^2_{likelihood}(\{\delta^i_j\})}{\partial \delta^i_m} &=&\sum_i \frac{ D^2\,A_i}{\bar{N}} \sum_p \vec{W}'(\vec{x}_p -\vec{x}_i) \left( \,\frac{\partial \vec{K}^2_p(\delta^i)}{\partial \delta^i_m} \right)\, , \nonumber \\
&=& \sum_p \sum_i \frac{  D^2\,A_i}{\bar{N}}  \vec{W}'(\vec{x}_p -\vec{x}_i)  \sum_j \vec{V}'(\vec{q}_p-\vec{x}_j) \frac{\Phi^2_j}{\partial \delta^i_m}  \nonumber \\
&=&   D^2 \sum_j F_j \frac{\Phi^2_j}{\partial \delta^i_m} \, ,  \nonumber \\
\end{eqnarray}
The second order Lagrangian potential \(\Phi^2_j\) can be calculated as
\begin{eqnarray}
\label{eq:Phi_LPT2}
\Phi^2_j &=& \sum_k \frac{-1}{k_k^2} {\rm{e}}^{2\pi j\,k \frac{\sqrt{-1}}{N}}  \sum_n \phi_n {\rm{e}}^{-2\pi n\,k \frac{\sqrt{-1}}{N}}  \, . 
\end{eqnarray}
with \(\phi_n\) given as:

\begin{eqnarray}
\label{eq:Phi_LPT2_a}
\phi_n  &=& \sum_{a>b} \phi^{aa}_n\,\phi^{bb}_n - \left(\phi^{ab}_n\right)^2 \, , 
\end{eqnarray}
where the individual potentials \(\phi^{ab}_n\) are related to the signal \(\delta^i_n\) via
\begin{eqnarray}
\label{eq:Phi_ZA}
\phi^{ab}_n &=& \sum_k \frac{k^a_k\,k^b_k}{k_k^2} {\rm{e}}^{2\pi n\,k \frac{\sqrt{-1}}{N}}  \sum_l \delta^i_l {\rm{e}}^{-2\pi l\,k \frac{\sqrt{-1}}{N}}  \, . 
\end{eqnarray}

With these definitions, we can write

\begin{eqnarray}
\label{eq:LH_Forces_d2}
\frac{\partial\psi^2_{LH}(\delta^i)}{\partial \delta^i_m} &=& D^2 \sum_j F_j \sum_k \frac{-1}{k_k^2} {\rm{e}}^{2\pi j\,k \frac{\sqrt{-1}}{N}}  \sum_n \frac{\partial \phi_n}{\partial \delta^i_m} {\rm{e}}^{-2\pi n\,k \frac{\sqrt{-1}}{N}} \, ,  \nonumber \\
&=& D^2\sum_j F_j \sum_k \frac{-1}{k_k^2} {\rm{e}}^{2\pi j\,k \frac{\sqrt{-1}}{N}}  \sum_n  {\rm{e}}^{-2\pi n\,k \frac{\sqrt{-1}}{N}} \nonumber \\
& & \,\frac{\partial}{\partial \delta^i_m} \left( \sum_{a>b} \phi^{aa}_n\,\phi^{bb}_n - \left(\phi^{ab}_n\right)^2\right)  \, ,  \nonumber \\
&=& D^2\sum_{a>b} \sum_j F_j \sum_k \frac{-1}{k_k^2} {\rm{e}}^{2\pi j\,k \frac{\sqrt{-1}}{N}}  \sum_n  {\rm{e}}^{-2\pi n\,k \frac{\sqrt{-1}}{N}} \nonumber \\
& & \,\frac{\partial}{\partial \delta^i_m} \left(  \phi^{aa}_n\,\phi^{bb}_n - \left(\phi^{ab}_n\right)^2\right)  \, ,  \nonumber \\
&=& D^2\sum_{a>b} \sum_j F_j \sum_k \frac{-1}{k_k^2} {\rm{e}}^{2\pi j\,k \frac{\sqrt{-1}}{N}}  \sum_n  {\rm{e}}^{-2\pi n\,k \frac{\sqrt{-1}}{N}} \nonumber \\
& & \, \left(  \frac{\partial \phi^{aa}_n }{\partial \delta^i_m}\,\phi^{bb}_n +\frac{\partial \phi^{bb}_n }{\partial \delta^i_m}\,\phi^{aa}_n - 2 \phi^{ab}_n \frac{\partial \phi^{ab}_n }{\partial \delta^i_m} \right)  \, ,  \nonumber \\
\end{eqnarray}
In the following we will discuss the individual terms. To simplify the notation, we introduce the tensor \(\tau^{abcd}\) defined as:
\begin{eqnarray}
\label{eq:LH_Forces_d2a}
\tau_m^{abcd}&=&\sum_j F_j \sum_k \frac{-1}{k_k^2} {\rm{e}}^{2\pi j\,k \frac{\sqrt{-1}}{N}}  \sum_n  {\rm{e}}^{-2\pi n\,k \frac{\sqrt{-1}}{N}}  \,  \frac{\partial \phi^{ab}_n }{\partial \delta^i_m}\,\phi^{cd}_n     \nonumber \\
&=& \sum_j F_j \sum_k \frac{-1}{k_k^2} {\rm{e}}^{2\pi j\,k \frac{\sqrt{-1}}{N}}  \sum_n  {\rm{e}}^{-2\pi n\,k \frac{\sqrt{-1}}{N}} \phi^{cd}_n \nonumber \\
& &  \sum_p \frac{k^a_p\,k^b_p}{k_p^2} {\rm{e}}^{2\pi n\,p \frac{\sqrt{-1}}{N}}  \sum_l \delta^K_{lm} {\rm{e}}^{-2\pi l\,p \frac{\sqrt{-1}}{N}} \, \nonumber \\
&=& \sum_j F_j \sum_k \frac{-1}{k_k^2} {\rm{e}}^{2\pi j\,k \frac{\sqrt{-1}}{N}}  \sum_n  {\rm{e}}^{-2\pi n\,k \frac{\sqrt{-1}}{N}} \phi^{cd}_n \nonumber \\
& &  \sum_p \frac{k^a_p\,k^b_p}{k_p^2} {\rm{e}}^{2\pi n\,p \frac{\sqrt{-1}}{N}}  {\rm{e}}^{-2\pi m\,p \frac{\sqrt{-1}}{N}} \, \nonumber \\
&=&  \sum_p \frac{k^a_p\,k^b_p}{k_p^2} {\rm{e}}^{-2\pi m\,p \frac{\sqrt{-1}}{N}} \sum_n {\rm{e}}^{2\pi n\,p \frac{\sqrt{-1}}{N}} \phi^{cd}_n \nonumber \\
& &  \sum_k  {\rm{e}}^{-2\pi n\,k \frac{\sqrt{-1}}{N}}\,\frac{-1}{k_k^2} \sum_j F_j {\rm{e}}^{2\pi j\,k \frac{\sqrt{-1}}{N}}\nonumber \\
&=&  \sum_p \frac{k^a_p\,k^b_p}{k_p^2} {\rm{e}}^{-2\pi m\,p \frac{\sqrt{-1}}{N}} \sum_n {\rm{e}}^{2\pi n\,p \frac{\sqrt{-1}}{N}} \phi^{cd}_n\, J_n \nonumber \\
\end{eqnarray}
With these definitions the second order Lagrangian contribution to the Hamiltonian force can be calculated as
\begin{eqnarray}
\label{eq:LH_Forces_e2}
\frac{\partial\psi^2_{LH}(s)}{\partial \delta^i_m} &=& D^2 \sum_{a>b} \left( \tau_m^{aabb} + \tau_m^{bbaa} - 2 \tau_m^{abab} \right) \, ,  \nonumber \\
\end{eqnarray}

This finally yields the Hamiltonian forces corresponding to the likelihood term:
\begin{eqnarray}
\label{eq:LH_Forces_final}
\frac{\partial\Psi_{likelihood}(\{\delta^i_j\})}{\partial \delta^i_m}  &=& - D^1\, J_m + D^2 \sum_{a>b} \left( \tau^{aabb} + \tau^{bbaa} - 2 \tau^{abab} \right) \, .\nonumber \\
\end{eqnarray}

\section{Adjoint FFT}
\label{USE_FFTS}
The following operation can be performed via FFT methods, when accounting  for adjoining the operation:
\begin{eqnarray}
\label{eq:a}
\sum_j a_j\,  {\rm{e}}^{2\pi j\,k \frac{\sqrt{-1}}{N}} &=& \sum_j \sum_q \hat{a}_q\,{\rm{e}}^{2\pi j\,q \frac{\sqrt{-1}}{N}}\,  {\rm{e}}^{2\pi j\,k \frac{\sqrt{-1}}{N}} \nonumber \\
&=& \sum_q \hat{a}_q\, \sum_j {\rm{e}}^{2\pi j\,(q+k) \frac{\sqrt{-1}}{N}} \nonumber \\
&=& \sum_q \hat{a}_q\, \delta^K{q,-k} \nonumber \\
&=& \hat{a}_{-k}\, \nonumber \\
&=& \hat{a}^{*}_{k}\, ,
\end{eqnarray}
where we made use of the fact that \(a_j\) is a real quantity, and the \(*\) denotes complex conjugation. Therefore, equation (\ref{eq:a}) simply describes the application of an FFT followed by a complex conjugation. To solve the adjoint Poisson equation we calculate:
\begin{eqnarray}
\label{eq:b}
\sum_k \frac{\hat{a}^{*}_{k}}{k_k^2} {\rm{e}}^{-2\pi m\,k \frac{\sqrt{-1}}{N}} &=& \sum_k \hat{b}_{k} {\rm{e}}^{-2\pi m\,k \frac{\sqrt{-1}}{N}} \nonumber \\
&=& \sum_k  \sum_j b_{j} {\rm{e}}^{-2\pi j\,k \frac{\sqrt{-1}}{N}}  {\rm{e}}^{-2\pi m\,k \frac{\sqrt{-1}}{N}} \nonumber \\
&=& \sum_j b_{j}  \sum_k {\rm{e}}^{-2\pi (j+m)\,k \frac{\sqrt{-1}}{N}}  \nonumber \\
&=& \sum_j b_{j}  \delta^K_{j,-m}  \nonumber \\
&=& b_{-m}  \nonumber \\
&=& b_{N-m}  \, ,
\end{eqnarray}
where in the last step we made use of the periodicity of the signal.

\section{Hamiltonian Masses}
\label{HAMILTONIAN_MASS}
A good guess for the Hamiltonian masses can greatly improve the efficiency of the hybrid Hamiltonian sampler. In order to derive appropriate Hamiltonian masses for the 2LPT-Poissonian system we will follow a similar approach as described in \citet{TAYLOR2008} and \citet{JASCHE2010HADESMETHOD}.
Since the efficiency of the Hamiltonian sampler depends on the accuracy of the leapfrog scheme, we will perform an approximated stability analysis of the integrator. The goal of this analysis is to find an expression for the Hamiltonian masses which optimizes the stability of the integration scheme for the 2LPT-Poissonian system.

According to the leapfrog scheme, given in equations (\ref{eq:LEAPFROG1}), (\ref{eq:LEAPFROG2}) and (\ref{eq:LEAPFROG3}), a single application of the leapfrog method can be written in the form:
\begin{equation}
\label{eqn:LEAPFROG_STEP_1_full}
p_m(t+\epsilon)=p_m(t) -\frac{\epsilon}{2} \left ( \left .\frac{\partial \Psi(\delta^i)}{\partial \delta^i_i}\right|_{\delta^i(t)} + \left .\frac{\partial \Psi(\delta^i)}{\partial \delta^i_m}\right|_{\delta^i(t+\epsilon)} \right)
\end{equation}

\begin{equation}
\label{eqn:LEAPFROG_STEP_2_full}
s_m(t+\epsilon)=s_m(t) +\epsilon \sum_j M^{-1}_{mj}\, p_j(t) -\frac{\epsilon^2}{2} \sum_j M^{-1}_{mj} \left .\frac{\partial \Psi(\delta^i)}{\partial \delta^i_j}\right|_{\delta^i(t)} \, . 
\end{equation}
We will then expand the Hamiltonian forces given in equation (\ref{eq:Hamiltonian_forces}) around a fixed value \((\delta^i)^0_m\), which is assumed to be the mean signal around which the sampler will oscillate once it left the Burn-in phase. Further, we will only expand up to linear order in the forces, which amounts to second order in the potential and hence to a Gaussian approximation of the 2LPT-Poissonian posterior distribution. For simplicity we will also ignore the second order Lagrangian term in the forces.
Thus, the Hamiltonian forces can be written as:
\begin{eqnarray}
\label{eq:APPROX_FORCES}
\frac{\partial\Psi(\{\delta^i_j\})}{\partial \delta^i_m} &=& \frac{\partial\Psi_{prior}(\{\delta^i_i\})}{\partial \delta^i_m} + \frac{\partial\Psi_{likelihood}(\{\delta^i_i\})}{\partial \delta^i_m}  \nonumber \\ 
&=& \sum_{j} S^{-1}_{mj}\,\delta^i_j - D^1\, J_m  \nonumber \\
& \approx & \sum_{j} S^{-1}_{mj}\,\delta^i_j - D^1\left(J_m((\delta^i)^0) + \left. \frac{\partial J_m(\delta^i)}{\partial \delta^i_m}\right|_{\delta^i_m=(\delta^i)^0_m} \, (\delta^i_m-(\delta^i)^0_m)\right)  \nonumber \\
& = & \sum_{j} \left (S^{-1}_{mj} -\delta^K_{mj}\, D^1\,\left. \frac{\partial J_j(\delta^i)}{\partial \delta^i_j}\right|_{\delta^i_j=(\delta^i)^0_j} \right )\,\delta^i_j \nonumber \\
&  & - D^1\left(J_m((\delta^i)^0) - \left. \frac{\partial J_m(\delta^i)}{\partial \delta^i_m}\right|_{\delta^i_m=(\delta^i)^0_m} \, (\delta^i)^0_m\right) \, . \nonumber \\
\end{eqnarray}
We will simplify the notation by introducing the matrix:
\begin{equation}
\label{eqn:A_MATRIX}
A_{mj}= S^{-1}_{mj} -\delta^K_{mj}\, D^1\,\left. \frac{\partial J_j(\delta^i)}{\partial \delta^i_j}\right|_{\delta^i_j=(\delta^i)^0_j}
\end{equation}
and the vector:
\begin{equation}
\label{eqn:D_VECTOR}
D_{m}= - D^1\left(J_m((\delta^i)^0) - \left. \frac{\partial J_m(\delta^i)}{\partial s_m}\right|_{\delta^i_m=(\delta^i)^0_m} \, (\delta^i)^0_m\right) \, .
\end{equation}
Equation (\ref{eq:APPROX_FORCES}) can then be written as:
\begin{eqnarray}
\label{eq:APPROX_FORCES_A}
\frac{\partial\Psi(\{\delta^i_j\})}{\partial \delta^i_m} &=& \sum_{j} A_{mj}\,\delta^i_j+D_m \, .
\end{eqnarray}
Introducing this approximation into equations (\ref{eqn:LEAPFROG_STEP_1_full}) and (\ref{eqn:LEAPFROG_STEP_2_full}) yields: 

\begin{eqnarray}
\label{eqn:LEAPFROG_STEP_1_full_a}
p_i(t+\epsilon)&=&\sum_{m} \left[\delta^{K}_{im}-\frac{\epsilon^2}{2} \sum_j A_{ij} M^{-1}_{jm}\right]\, p_m(t) \nonumber \\
& &-\epsilon\sum_{j} A_{ij} \sum_p\left[\delta^{K}_{jp}-\frac{\epsilon^2}{4}\sum_m M^{-1}_{jm}\,A_{mp}\right]\,r_p(t)\nonumber \\
& & - \frac{\epsilon}{2} \sum_m \left [ \delta^{K}_{im} - \frac{\epsilon^2}{2} \sum_j A_{ij} M^{-1}_{jm} \right ] D_m\nonumber \\
\end{eqnarray}
and
\begin{eqnarray}
\label{eqn:LEAPFROG_STEP_2_full_a}
r_i(t+\epsilon)&=&\epsilon \sum_j M^{-1}_{ij} \, p_j(t) \nonumber \\
& &+\sum_{m}\left(\delta^K_{im} -\frac{\epsilon^2}{2}\sum_j M^{-1}_{ij} \,A_{jm}\right)r_m(t)\nonumber \\
& & -\frac{\epsilon^2}{2}\sum_j M^{-1}_{ij}D_j\, . \nonumber \\ 
\end{eqnarray}
This result can be rewritten in matrix notation as:
\begin{equation}
\label{eqn:vector}
\left (\begin{array}{c} r(t+\epsilon)\\ p(t+\epsilon)\end{array}\right ) = T \left (\begin{array}{c} r(t)\\ p(t)\end{array}\right ) - \frac{\epsilon^2 }{2}\left(\begin{array}{c} M^{-1}\,D\\ \epsilon \left [ \rm{I} -\frac{\epsilon^2}{2}A\,M^{-1}\right] D\end{array}\right )  \, ,
\end{equation}
where the matrix \(T\) is given as:
\begin{equation}
\label{eqn:vector}
T=\left (\begin{array}{cc} \left [\rm{I}-\frac{\epsilon^2}{2} M^{-1} A\right] &  \epsilon M^{-1}\\ -\epsilon\,A\left[\rm{I}-\frac{\epsilon^2}{4}M^{-1}\,A\right] & \left[\rm{I}-\frac{\epsilon^2}{2} A\,M^{-1}\right]\end{array}\right ) \, ,
\end{equation}
with \(\rm{I}\) being the identity matrix.
Successive applications of the leapfrog step yield the following propagation equation:
\begin{equation}
\label{eqn:iterative_vector}
\left (\begin{array}{c} r^n\\ p^n\end{array}\right ) = T^n \left (\begin{array}{c} r^0\\ p^0\end{array}\right ) - \frac{\epsilon^2 }{2} \left [\sum^{n-1}_{i=0} T^i \right] \left(\begin{array}{c} M^{-1}\,D\\ \epsilon \left [ \rm{I} -\frac{\epsilon^2}{2}A\,M^{-1}\right] D\end{array}\right )  \, .
\end{equation}
This equation demonstrates that  there are two criteria to be fulfilled if the method is to be stable under repeated application of the leapfrog step.
First we have to ensure that  the first term of equation (\ref{eqn:iterative_vector}) does not diverge. This can be fulfilled if the eigenvalues of \(T\) have unit modulus. The eigenvalues \(\lambda\) are found by solving the characteristic equation:
\begin{equation}
\label{eqn:characteristic_equation}
det\left[ \rm{I}\, \lambda^2 - 2\,\lambda \left ( \rm{I} -\frac{\epsilon^2}{2}A\,M^{-1}\right) + \rm{I} \right ]=0 \, .
\end{equation}
Note that this is a similar result to what was found in \citet{TAYLOR2008}.
Our aim is to explore the parameter space rapidly, and therefore we wish to choose the largest \(\epsilon\) still compatible with the stability criterion. However, any dependence of equation (\ref{eqn:characteristic_equation}) also implies that no single value of \(\epsilon\) will ensure unit modulus for every eigenvalue. For this reason we choose:
\begin{equation}
\label{eqn:stability_condition}
A=M \, .
\end{equation}
We then obtain the characteristic equation:
\begin{equation}
\label{eqn:characteristic_equation_a}
\left[ \lambda^2 - 2\,\lambda \left ( 1 -\frac{\epsilon^2}{2}\right) + 1 \right ]^N=0 \, ,
\end{equation}
where \(N\) is the number of voxels. This yields the eigenvalues:
\begin{equation}
\label{eqn:eigenvals}
\lambda = \pm \, i \sqrt{1-\left[1 -\frac{\epsilon^2}{2} \right]^2} +\left[1 -\frac{\epsilon^2}{2} \right] \, ,
\end{equation}
which have unit modulus for \(\epsilon \le 2 \).
The second term in equation (\ref{eqn:iterative_vector}) involves evaluation of the geometric series \(\sum^{n-1}_{i=0} T^i\).
The geometric series for a matrix converges if and only if \(| \lambda _i| < 1 \) for each \( \lambda _i \) eigenvalue of \(T\). 
This clarifies that the nonlinearities in the Hamiltonian equations generally do not allow for arbitrary large pseudo time steps \(\epsilon\). In addition, for practical purposes we usually restrict the mass matrix to the diagonal of equation (\ref{eqn:A_MATRIX}).
In practice we choose the pseudo time step \(\epsilon\) as large as possible while still obtaining a reasonable rejection rate.

Given these assumptions we can assume the mass matrix to be:
\begin{equation}
\label{eqn:MASS_MATRIX}
M_{mj}= S^{-1}_{mj} -\delta^K_{mj}\, D^1\,\left. \frac{\partial J_j(\delta^i)}{\partial \delta^i_j}\right|_{\delta^i_j=(\delta^i)^0_j}\, ,
\end{equation}
where 
\begin{eqnarray}
\label{eqn:MASS_MATRIX}
\frac{\partial J_m(s)}{\partial s_m} &=& \sum_k \frac{-1}{k_k^2} {\rm{e}}^{-2\pi m\,k \frac{\sqrt{-1}}{N}} \sum_j \frac{\partial F_j}{\partial \delta^i_m}\,  {\rm{e}}^{2\pi j\,k \frac{\sqrt{-1}}{N}} \, \nonumber \\
&=& \sum_k \frac{-1}{k_k^2} {\rm{e}}^{-2\pi m\,k \frac{\sqrt{-1}}{N}} \sum_j {\rm{e}}^{2\pi j\,k \frac{\sqrt{-1}}{N}} \,\nonumber \\
& & \sum_p \sum_i \left ( \frac{1}{\bar{N}}  \vec{W}'(\vec{x}_p -\vec{x}_i)   \vec{V}'(\vec{q}_p-\vec{x}_j)\frac{\partial A_i}{\partial \delta^i_m} \right.\nonumber \\
& &  \left . \frac{1}{\bar{N}}  \vec{W}''(\vec{x}_p -\vec{x}_i)   \vec{V}'(\vec{q}_p-\vec{x}_j)\, A_i \frac{\partial x_p}{\partial \delta^i_m} \right)\, , \nonumber \\
\end{eqnarray}
where we used of equations (\ref{eq:Fj}) and (\ref{eq:JM}). 
According to equation (\ref{eq:Ai}) \(\frac{\partial A_i}{\partial \delta^i_m}\) can be expressed as:
\begin{eqnarray}
\label{eqn:MASS_MATRIX}
\frac{\partial A_i}{\partial s_m} &=&  \frac{ R_i\, \bar{N}}{\left(R_i\, \bar{N}\,(1+G(a,\delta^i)_i)\right)^2} \frac{\partial \delta_i(\delta^i)}{\partial \delta^i_m}  \, . \nonumber \\
&=& B_i\, \frac{\partial G(a,\delta^i)_i}{\partial \delta^i_m}\, ,
\end{eqnarray}
where we introduced the quantity \(B_i=  \left( R_i\, \bar{N} \right) / \left(R_i\, \bar{N}\,(1+G(a,\delta^i)_i)\right)^2\) to simplify notation.
We then arrive at the expression:
\begin{eqnarray}
\label{eqn:MASS_MATRIX}
\frac{\partial J_m(\delta^i)}{\partial \delta^i_m} &=& \sum_k \frac{-1}{k_k^2} {\rm{e}}^{-2\pi m\,k \frac{\sqrt{-1}}{N}} \sum_j \frac{\partial F_j}{\partial \delta^i_m}\,  {\rm{e}}^{2\pi j\,k \frac{\sqrt{-1}}{N}} \, \nonumber \\
&=&  \sum_i \sum_k \frac{-1}{k_k^2} {\rm{e}}^{-2\pi m\,k \frac{\sqrt{-1}}{N}} \sum_j {\rm{e}}^{2\pi j\,k \frac{\sqrt{-1}}{N}} \,\nonumber \\
& & \sum_p  \frac{1}{\bar{N}}  \vec{W}'(\vec{x}_p -\vec{x}_i)   \vec{V}'(\vec{q}_p-\vec{x}_j)\,B_i\, \frac{\partial G(a,\delta^i)_i}{\partial \delta^i_m} \, , \nonumber \\
\end{eqnarray}

\bsp
\label{lastpage}

\end{document}